\documentclass{aa}  

\usepackage{graphicx}
\usepackage{txfonts}
\usepackage{subcaption}
\usepackage{longtable}
\usepackage{lscape}
\usepackage{placeins}

\usepackage{xcolor}
\definecolor{blue} {rgb} {0.1,0.1,0.8}
\definecolor{green} {rgb} {0.0,0.6,0.0}
\definecolor{magenta} {rgb} {0.6,0.1,0.6}

\begin{document}

   \title{Spatio-temporal progression of star formation in the \mbox{Carina~OB1 association}}

   \author{C. G{\"o}ppl
          \inst{1,2}
          \and
          T. Preibisch\inst{1}
          }

   \institute{Universit\"ats-Sternwarte M\"unchen, 
              Ludwig-Maximilians-Universit\"at,
              Scheinerstr.~1, 81679 M\"unchen, Germany
              \and
              Excellence Cluster ORIGINS, Boltzmannstr.~2, 85748 Garching, Germany\\
              \email{cgoeppl@usm.uni-muenchen.de}
             }

\titlerunning{Star Formation History of the Car~OB1 Association}
\authorrunning{G{\"o}ppl \& Preibisch}

   \date{Received 20 November 2025; accepted 12 September 2026}

  \abstract
   {The Carina~OB1 association, harboring the famous Carina Nebula near its center, is the most massive, well-studied OB association in our Galaxy. Containing several massive star clusters with different ages, it shows a complex spatio-temporal structure.}
   {We explore the formation history and kinematic evolution of the eight most massive clusters in Car~OB1.}
   {We determine the cluster ages, stellar populations, the initial massive star content, and the number of supernova explosions since the time of each cluster’s formation. Using \textit{Gaia}~DR3 astrometry, we determine the 3D configuration of the association and perform a traceback analysis of the cluster positions over the last $ \approx 13$~Myr.}
   {The traceback analysis places the birth location of the oldest ($ \simeq 13$\;Myr) cluster NGC~3293 in the Carina spiral arm. Approximately four Myr later, a series of
   $\simeq 11$ supernova explosions started in NGC~3293.
 We construct a possible scenario in which the expanding bubble driven by these supernova explosions interacted with the southern part of the original cloud complex and led to the sequential formation of the younger clusters in the central Carina Nebula. 
   }
   {Our scenario suggests that the sequence of $ \approx 16 \pm 4$ supernova explosions in the association during the last 10\;Myr played an important role in creating the massive clusters in Car~OB1. However, the supernova feedback does not dominate the large-scale expansion of the association; the general expansion rather seems to reflect the initial kinematics of the original unbound cloud complex.}  
\maketitle

\section{Introduction}

OB associations are the stellar nurseries where most of the high-mass stars in our Galaxy are born \citep[see, e.g.,][for a recent review]{2023ASPC..534..129W}. Due to their large spatial extent (several tens of parsecs) and low stellar density, associations are expected to expand and disperse over a timescale of a few tens of megayears. Most OB associations show substructure in the form of stellar groups and clusters of various ages; many OB associations also still contain molecular clouds with ongoing star formation. This is seen, for example, in the Scorpius-Centaurus (Sco-Cen) association \citep{2019A&A...623A.112D, 2023A&A...678A..71R,2023A&A...677A..59R, 2023MNRAS.519.3992Z} and in Orion OB1 \citep{2019A&A...628A.123Z,2020A&A...643A.114C,2023A&A...673A.114H}.
Indications for sequential, presumably triggered formation of subgroups or clusters in OB associations were found decades ago in several nearby OB associations \citep[see, e.g.,][]{1977ApJ...214..725E,1992A&A...262..258D,1999AJ....117.2381P,2007prpl.conf..345B}.

In the era before \textit{Gaia}, model scenarios for triggered sequential star formation in OB associations were necessarily rather simple, usually employing only a few steps of triggering in a linear sequence of large star clusters and/or groups.
In the last few years, the availability of precise astrometric data from \textit{Gaia} allowed a much more reliable definition of clusters and associations as co-moving groups of stars. For nearby OB associations, the precise \textit{Gaia} proper motions and parallaxes made it possible to draw a detailed picture of their spatio-temporal evolution. In some cases, these studies revealed a clear systematic spatial progression of star formation events, creating chains of clusters with decreasing ages \citep[e.g.,][in the case of the Sco-Cen OB association]{2023A&A...677A..59R,2023A&A...679L..10P, 2025A&A...693A.175P, 2025A&A...694A..60M, 2026A&A...709A.181G, 2026A&A...705A.108H}, which connect the oldest cluster with very young clusters, which are sometimes still sites of ongoing star formation.
Similar, although often less clear, indications of sequential star formation have also been found in several other OB associations, such as Cepheus-Hercules \citep{2024ApJ...975...99K}, the Circinus Complex \citep{2025ApJ...985..111K}, and the Vela OB2 Complex \citep{2019A&A...621A.115C,2021ApJ...923...20P}. Sequential star formation is suggested to be triggered by feedback from massive stars, initiating the formation of the next generation of clusters, for example, via their (ionizing) radiation, winds, or supernova shocks.

\begin{figure}
    \centering
\resizebox{7.5cm}{!}{\includegraphics[trim=6.3cm 9.5cm 3.8cm 10.5cm, clip]{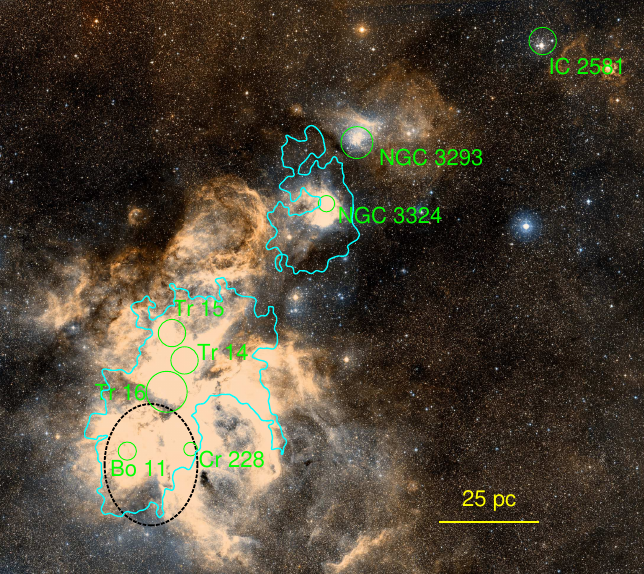}}
\resizebox{7.5cm}{!}{\includegraphics{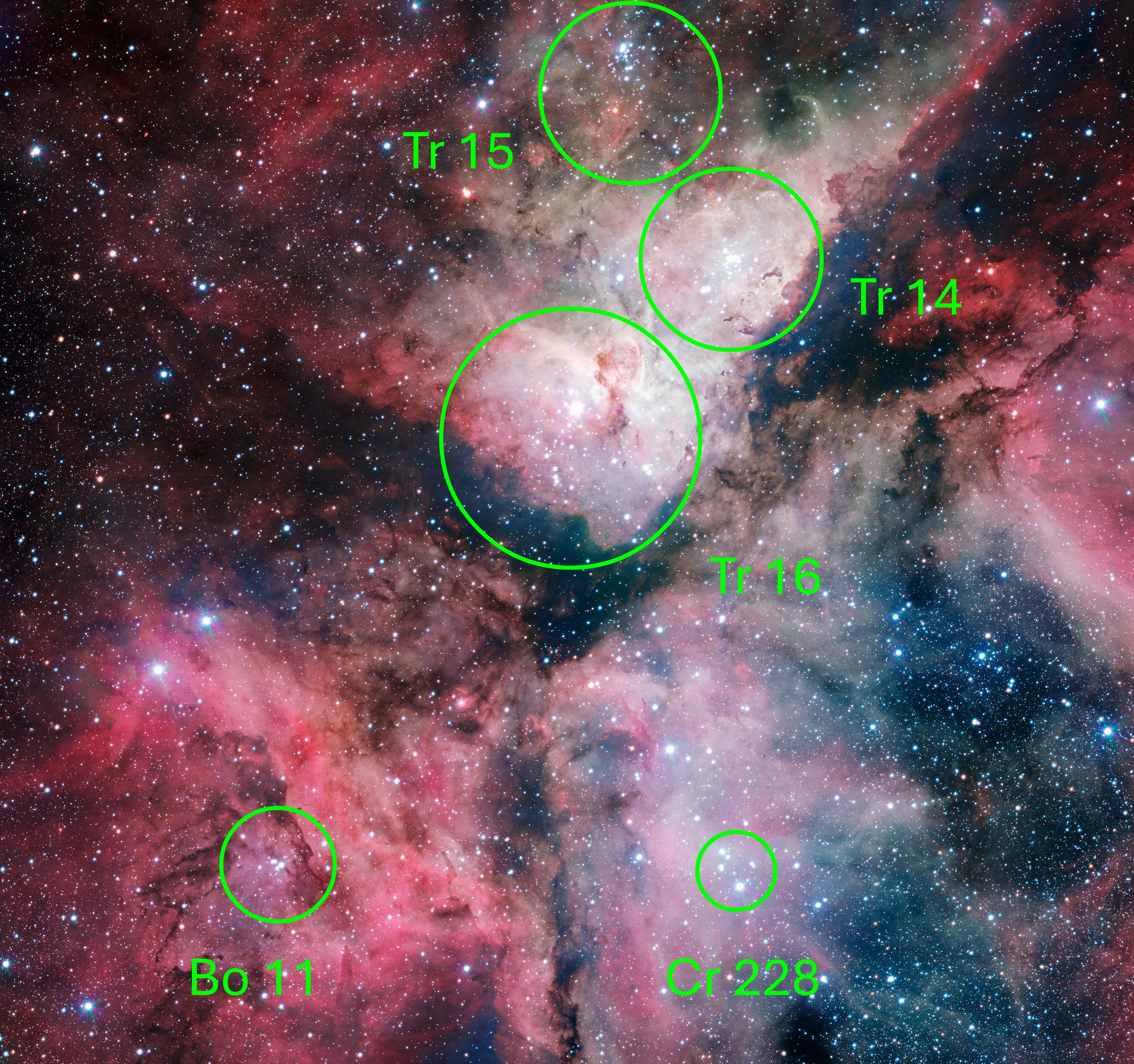}}
    \caption{Optical wide-field image of the full Car~OB1 association (top; Digitized Sky Survey~2, image credit: NASA/ESA/AAO, UK Schmidt Telescope)  and optical VST image of the central Carina Nebula (bottom; from: {\tt www.eso.org/public/images/eso1250a/}; image credit: ESO/VPHAS+ Consortium/Cambridge Astronomical Survey Unit).
  The main clusters are marked as annotated circles. The blue contour lines in the upper image mark the two areas with dense clouds, defined by a surface brightness of $\ge 10.9\;\rm mJy/(arcsec)^2$ in the \textit{Herschel} $160\;\mu \rm m$ map \citep{{2017ASPC..512..379G}}. The black dashed ellipse highlights the Southern Pillars.}
    \label{fig:CarOB1Clusters}
\end{figure}

The Carina~OB1 (Car~OB1) association is the most massive of the well-studied OB associations in our Galaxy, with a total population of $\approx 10^{5}$ stars \citep{1980AJ.....85.1193T, CCCP-Clusters,2019MNRAS.490..440L,2019ApJ...870...32K, 2025A&A...695A..48G}. 
Car~OB1 extends over about four degrees on the sky (see Fig.~\ref{fig:CarOB1Clusters}); at a distance of 2.35\;kpc \citep{2022A&A...660A..11G} this corresponds to a physical length scale of approximately 160\;pc. The association hosts at least 135 O-type, Wolf-Rayet, and supergiant stars \citep{1984A&AS...58..163L,2011ApJS..194....7N, 2025A&A...695A..48G}, many of which are concentrated in one of several stellar clusters, including the famous clusters Trumpler~14 and Trumpler~16 in the central Carina Nebula, and NGC~3293 in the northwestern part of Car~OB1. The association also still contains huge amounts of gas and molecular clouds with an estimated total mass of $\sim 10^6~M_{\odot}$ \citep{SB07, Preibisch12,2016MNRAS.456.2406R}. 

The visually most prominent part of Car~OB1 is the Carina Nebula \citep[see][]{SB08}, a bright \ion{H}{II} region illuminated by the numerous O-type stars in the central young clusters Tr\;14 \citep[$\approx 1$\;Myr,][]{HAWKI-survey,2024A&A...685A.100I} and Tr\;16 \citep[$\approx 3-4$\;Myr,][]{HAWKI-survey}. Just south of Tr\;16, very active star formation is still going on in the ``Southern Pillars'' region \citep[see, e.g.,][]{Smith10b,Povich11,Gaczkowski13}. 
However, the association also contains several slightly older clusters, such as Trumpler\;15 \citep[$ 5-8$\;Myr,][]{CNC-Laboca} in the central Carina Nebula and NGC\;3293 \citep[$ \approx  10-15$\;Myr,][]{2025A&A...695A..48G} in the northwestern part of Car~OB1.

In a previous study \citep[][GP25 hereafter]{2025A&A...695A..48G}, we used \textit{Gaia}~DR3 astrometric data \citep{Gaia, Gaia-DR3} for an unbiased search for and definition of stellar clusters in Car~OB1, and to determine their properties such as distance, radius, and internal kinematics. Figure~\ref{fig:CarOB1Clusters} shows the locations of the eight most massive clusters that qualified as members of the Car~OB1 association. We also compiled a catalog of OB stars in the association based on the literature and searched for new OB candidates using the effective temperatures provided by \textit{Gaia}~DR3.
In GP25, we identified numerous new O-star candidates in Car OB1
and found that a significant fraction of the O-type stars in Car OB1 cannot be clearly
assigned as a member of one of the clusters, since they are found to be widely
distributed across the full extent of the association.

A scenario for the recent formation history of the clusters Tr~14, 15, and 16 and the South Pillars clouds in the central Carina Nebula Complex (CNC) has been presented by \citet{Povich19}, who found that star formation activity in the CNC during the last 10 Myr showed peaks at the formation of Tr~15 $\approx 6.5$~Myr ago and in the last $\approx 3$~Myr with the birth of Tr~16 and 14.
In the study presented here, we extend the star formation scenario in two ways: firstly, we consider the full spatial extent of the Car~OB1 association, including also the northern parts with the clusters NGC~3324, NGC~3293, and IC~2581. Secondly, our scenario covers the full history of the association, which started $ \approx 13$\;Myr ago with the formation of the oldest cluster, NGC~3293.

As the first step of the analysis in this paper, we build on our previous results from GP25 to characterize the stellar populations of the clusters in Car~OB1 and estimate the number of supernova explosions that have occurred until today (Sect.~2).  
In Sect.~3, we determine the 3D structure and kinematics of the association, and
 perform a traceback analysis of the cluster positions over the formation history of Car~OB1.
In Sect.~4, we construct a model for the global formation history of the association, and we summarize our conclusions in Sect.~5.

\section{Data}

The stellar data for this study are based on the results of GP25,
complemented by eleven additional high-mass stars from \citet{2025MNRAS.543...63P}.
We consider here the eight most massive clusters in Car~OB1, selected under the condition that each cluster has at least two high-mass\footnote{For the conversion of (main-sequence) spectral type to stellar mass, we use the most recent version \url{https://www.pas.rochester.edu/~emamajek/EEM_dwarf_UBVIJHK_colors_Teff.txt} of the calibration table from \citet{2013ApJS..208....9P}, according to which all O-type main-sequence stars have a mass of $ M_\ast \ge 18\,M_\odot$.} ($ M_\ast \ge 18\,M_\odot$) stars today.
The other small clusters and stellar groups in Car~OB1 identified in GP25 are not included in the analysis in this paper because they have too few (in most cases $< 25$) members; this does not allow a meaningful determination of cluster ages and since none of these small clusters have available radial velocity data for more than two members, their 3D kinematics cannot be established in a meaningful way.

This results in an overall sample of 1\,625 stars, which all have \textit{Gaia}~DR3 counterparts within a 1~arcsecond radius, and 1573 have a full astrometric solution.  Basic data for all stars in each cluster are available as an electronic table at the CDS; a preview of the table is shown in Table~\ref{tab:CarOB1HighMass}.
We do not consider the possible substructure of the clusters, as the rather small numbers of subcluster members do not allow a reliable determination of their age or their 3D kinematics. 

We collected radial velocities of these stars, primarily from the
Gaia-ESO survey DR5 \citep{2023yCat..36760129H}, from \textit{Gaia} DR3 \citep{Katz2023}, and from various other surveys, as listed in Table~\ref{tab:RVAppendix}. The radial velocities are prioritized from the surveys in the order given in the table. We exclude radial velocities whose uncertainties are larger than $5\,\rm km\,s^{-1}$ or whose radial velocity differs by more than two standard deviations from the association’s median (see Fig.~\ref{fig:PMStars}). 
Velocities that are significantly different from the median may be caused by binaries, runaway stars, or further uncertainties. Therefore, they are probably not representative of the clusters’ motion. The combined quality criteria result in radial velocities for 398 stars.
For estimates of the full stellar population of the clusters, we also used the catalog of young stars emitting X-rays from the \textit{Chandra }Carina Complex Project \citep[CCCP,][]{CCCP-intro}, as well as the source catalogs resulting from the \textit{Chandra} X-ray observations of NGC~3324 \citep{Preibisch14} and NGC~3293 \citep{2017A&A...605A..85P}.

Cluster ages are compiled from our analysis in GP25 and a critical assessment of values published in the recent literature.
While for some of the clusters, age estimates are based on the analysis of
optical color-magnitude diagrams built on \textit{Gaia} photometry,
most of the younger clusters are still associated with (or embedded in) cloud structures of Car~OB1, and therefore suffer from considerable extinction, which hampers the interpretation of their
optical color-magnitude diagrams. Therefore, for these younger clusters literature values based on the analysis of near-infrared and X-ray data
are used.

\section{Methods}

\subsection{Estimates of the total cluster populations, cluster mass, and expected number of supernovae}
\label{sec:estimateclusterpop}

Information about the number of high-mass stars in a cluster is essential to quantify the level of feedback the cluster exerts on its surroundings.
For young ($\la 4$\;Myr old) clusters, where even the most massive stars are still present, one can simply count the number of observed O-type, WR (Wolf-Rayet) stars, and supergiants in the respective cluster area.

For the older clusters ($\ga 5$\;Myr), one also has to take into account that the most massive stars may have already exploded as supernovae. 
While there is no direct observational evidence of recent supernova explosions in Car~OB1 (e.g., in the form of a prominent supernova remnant), 
the substantial numbers of early B-type stars in the older ($\ge 5$\;Myr) clusters
NGC~3293, IC~2581, and Tr~15 strongly suggest that some O-type stars must have been initially present in these clusters. 
The X-ray detection of at least one
neutron star candidate in the Carina Nebula \citep{2009ApJ...695L...4H} provides direct support for this scenario.
Further indirect evidence comes from X-ray observations of the strong diffuse high-energy emission in the area around the Carina Nebula, which clearly show that the CNC superbubble is filled with large amounts of very hot X-ray-emitting plasma \citep{2011ApJS..194...15T,2024A&A...682A.172S}. Although a large fraction of this plasma is expected to be generated by the winds of numerous massive stars, it is considered quite likely that supernova explosions have contributed to the observed emission.
The fact that no prominent supernova remnant is seen in Car~OB1 does not contradict the expectation of several supernova explosions over the last $\approx 10$\;Myr, because 
the lifetime of a supernova remnant is rather short, $\leq 10^5 - 10^6$~years \citep[see, e.g.,][]{1989ApJ...340..355B,2022hxga.book...77B}.
The lack of prominent supernova signs in Car~OB1 thus implies that the last supernova explosions must have happened more than a few $10^5$ years ago.
For the three clusters older than 5\;Myr, we estimated the number of supernovae that are expected to have already occurred by using the Kroupa initial mass function (IMF) \citep{2001MNRAS.322..231K} and extrapolating based on the number of present OB stars, as described in more detail in Sect.~\ref{sec:clusternotes}.

We used two different approaches to estimate the total stellar population of each cluster.
The first approach is an IMF extrapolation based on the observed population of O-type (or B-type) stars in each cluster and extrapolating down to $0.08\,\rm M_{\odot}$. The upper mass limit was chosen for each cluster individually based on the most massive star, while the lower mass limit was chosen based on the present stellar population of each cluster. We take stars down to spectral type B3 to account for the extrapolations, as the overall high-mass star population in Car~OB1 can be assumed to be complete down to B3 (see Fig.~16 in GP25). We quantify the uncertainty of the extrapolated results by error propagation, assuming that the input values follow a Poisson distribution.

In the second approach, we estimated the total number of stellar members from the observed number of X-ray detected stars in each cluster, employing the number ratio of X-ray sources to stellar population in \citet{CCCP-Clusters} as they extrapolate a stellar population of 104\,000 stars from a X-ray population of 10\,728.
For clusters that were not recovered by cluster analysis in \citet{CCCP-Clusters}, for example, Collinder~228, we chose all stars detected by X-ray within the cluster radius (as listed in Table~\ref{tab:ClustersCarOB1Ages}) and extrapolated from that value. We also used this method for Tr~16, as our cluster boundary and the one used in \citet{CCCP-Clusters} deviate strongly from each other.

For most clusters, the two different estimates of the total stellar population agree within a factor of about two. The larger differences seen in some cases are related to the uncertainties of both methods, in particular to stochastic effects in the numbers of O-type stars, and illustrate the uncertainties of these extrapolations. Therefore, the star numbers should be considered estimates rather than precise values.

We then estimated the total stellar mass of the clusters by averaging the stellar populations from the high-mass star and X-ray data, and converting the determined average to stellar mass by assuming a mean stellar mass of $ 0.57\,\rm M_\odot$. 
For the conversion of (main-sequence) spectral type to stellar mass, we used the most recent version 
of the calibration table from \citet{2013ApJS..208....9P}. 
For evolved O-type stars, we converted the spectral type to stellar mass using the table provided by \citet{2010A&A...524A..98W}.
In the rest of the paper, the term ``high-mass star'' means a star with an initial main-sequence stellar mass of $ \ge 18\,M_\odot$.

\subsection{Timing of supernova explosions in Car OB1}
\label{sec:SN}

In order to construct a formation scenario for Car~OB1 that takes the effects of supernovae into account, we also need information about the times at which the first supernova explosion in each cluster occurred, since this dominates the dynamical evolution of the expanding bubble. 
As the most massive star in a cluster explodes first, we can estimate the time of the first supernova if we know the mass of the initially most massive star in the cluster. We estimated this maximum stellar mass in two ways: (i) by using the observed relation between the total mass of a cluster and the mass of the most massive star as given in \citet{2013MNRAS.434...84W}, and 
(ii) by extrapolating the IMF of the cluster based on the cluster mass as listed in Table~\ref{tab:ClustersCarOB1Ages} and calculating the mass bin $[M_{\ast, \rm max}, 100]~M_\odot$ in which we would only expect one star. 
From this estimate of the highest star mass, we can then determine its lifetime using the stellar tracks from \citet{2022A&A...658A.125S}. The results of this analysis are given in Table~\ref{tab:SNtime}. For NGC~3293, the first supernova should have occurred $\approx 9$~Myr ago. The first supernova in Tr~15 happened $\approx 2$~Myr ago. In total, there were $ \approx 16$ supernovae in Car~OB1 during the last $\approx 9$~Myr as listed in Table~\ref{tab:ClustersCarOB1Ages}. Since there is no evidence of a recent supernova in this area
\citep[see][]{2011ApJS..194...15T}, the last of these supernovae must have happened more than $ 10^5-10^6$~years ago.

\subsection{Kinematic traceback of the clusters}

The basis for our investigation of the 3D structure and kinematics of the clusters are the mean values for the positions, distances, proper motions, and radial velocities. We calculate these values and their uncertainties with stars from the Car~OB1 high-mass star catalog and the Car OB1 cluster member catalog (GP25) inside each cluster radius, using the maximum-likelihood method as described in \citet{2022A&A...660A..11G}.

In order to explore how the galactic location and the  spatial configuration of the clusters within the association have changed with time,
we utilize the galactic dynamics package \texttt{galpy} \citep{2015ApJS..216...29B} with the Milky Way's gravitational potential model\footnote{\texttt{MWPotential2014} models the three main gravitational components of the Milky~Way with a Miyamoto-Nagai potential for the disk \citep{1975PASJ...27..533M}, a power-law density spherical profile with an exponential cut-off for the bulge, and a Navarro-Frenk-White profile for the dark-matter halo \citep{1997ApJ...490..493N}.} \texttt{MWPotential2014} \citep{2015ApJS..216...29B} to trace back the position of clusters in Car~OB1 over the last 13~Myrs. 
For our analysis, we used a solar distance of $R_{\odot} = 8.23\pm0.12$~kpc \citep{2023MNRAS.519..948L} to the galactic center, a solar position of $z_{\odot}=17\pm5$~pc \citep{2017MNRAS.465..472K} above the galactic mid-plane, and solar motion of $(U_{\odot}, V_{\odot}, W_{\odot}) = \left(11.1^{+0.69}_{-0.75}, 12.24^{+0.47}_{-0.47}, 7.25^{+0.37}_{-0.36}\right)~\rm km\,s^{-1}$ \citep{2010MNRAS.403.1829S}. 
We performed the traceback 1000 times, adding errors randomly weighted by a normal distribution to each kinematic cluster variable, and determined the mean position and standard deviation of each cluster at time steps of 0.1~Myr. When the traceback time exceeds the age of a cluster, we assume to be tracing back the star's natal cloud, which is an approximation, as the 3D motions of gas clouds are influenced by their environment. We also traced back the expected location of the Carina spiral arm according to \citet{2019ApJ...885..131R}, assuming a galactic spiral pattern speed of $ \Omega_p=28.2\pm2.1\,\rm km\,s^{-1}\,kpc^{-1}$ \citep{2019MNRAS.486.5726D}.

\section{Results}

Table~\ref{tab:ClustersCarOB1Ages} lists the age, total stellar population and mass, number of high-mass stars and supernovae, and time of the first supernova for each cluster and the full Car~OB1 region. As the existing X-ray data do not fully cover the entire region, extrapolations based on high-mass stars are the only feasible approach for the full Car~OB1. Table~\ref{tab:ClusterVel} lists the calculated mean values for the positions, distances, proper motions, and radial velocities for the clusters. Table~\ref{tab:ClustersCarOB1Ages} lists the number of cluster members that were selected by DBSCAN in GP25; the number of cluster members in Table~\ref{tab:ClusterVel} is higher, as we also include the OB stars in a cluster that were not identified by DBSCAN for the analysis.

\begin{table*}
\caption{Stellar population properties of the eight largest clusters in Car~OB1.}
    \centering
    \resizebox{\hsize}{!}{\begin{tabular}{lclrcrrrrrr}
    \hline\hline
       Cluster & Radius & Age & $N_{\ge  18\;M_\odot}$ & $\mathcal{E}[N_{\rm SN}]$ &$t_{\rm first\; SN}$ & $\mathcal{E}[N_{\ge  18\;M_\odot}]$ & $\mathcal{E}[N_{\rm tot}]$ & $\mathcal{E}[N_{\rm tot}]$ & $ \mathcal{E}[M_{\rm tot}]$ & $ N_{\rm DBSCAN}$\\
        &  (pc) & (Myr)&  today & & (Myr)  & initial & X-ray & O/B &  $ M_\odot$\\
        \hline
        Tr 16 & 5.1& $\simeq \, 3\!-\!4$  & 35 & 0 & -- & 35 & $ 11\,900\pm400$ &  $ 18\,300 \pm 3100$ & $ 8700\pm1800$ &  131\\   
        Tr 14 & 3.4 &$\approx  \; 1 \; (^{+3}_{-1})$  & 18 & 0 & -- & 18 & $ 14\,000\pm400$ & $ 9400\pm2300$ & $ 6700\pm1300$ & 448\\  
        Tr 15 & 3.4&$\approx \; 6 \; (\pm2)$  & 4 &  $ 2 \pm 1$ & $\approx -2$ & 6 & $ 4700\pm300$ & $ 2900\pm 700$ & $ 2200\pm400$ &  154\\ 
        Cr 228 & 1.7& $\approx \; 4 \; (\pm1)$ & 5 & 0 & -- & 5 & $  600\pm 100$ &  $ 2800\pm1300$ & $ 1000\pm800$ & 40\\
        Bo 11 &2.3&$\approx \; 4\!-\!5$  & 4 & 0 & -- & 4 & $ 1300\pm200$ &  $  2400\pm1300$ & $ 1100\pm700$ &  46\\
        NGC 3324  &2.1&$\approx  \; 2 \; (\pm1)$& 3 &  0  & -- & 3 & $  1600\pm200$ & $ 2500\pm1500$ & $ 1200\pm900$ &  83 \\
        NGC 3293  &3.9& $ \approx 13 \; (\pm 2)$  & 3 & $ 11 \pm 3$ & $\approx -9$ &11&  $ 5400\pm300$ & $ 5200\pm800$ & $ 3000\pm500$ &  358\\ 
        IC 2581  &3.5&$\approx 12$  & 2 &  $ 3 \pm 2$ & $ \approx -6$ & 4 &  -- &$ 1800\pm700$ & $ 1000\pm400$ &  119\\     
        \hline \hline
        \multicolumn{3}{l}{ Distributed population} & 61 & & &> 61 &&$ \ga 30\,000$& & \\
         \multicolumn{3}{l}{Full Car~OB1} & $\ge 135$ & $ 16\pm 4$& & $ \ga 147$  &  & $\ga 75\,000$ & & \\
         \hline
    \end{tabular} }
    \tablefoot{The clusters are listed with their radius (from GP25), their age, the number of currently observed high-mass ($M \ge 18\;M_\odot$) stars, our estimates for the expected number $\mathcal{E}[N_{\rm SN}]$ of supernovae that happened in each cluster (Sect.~\ref{sec:estimateclusterpop}), the number of expected initial high-mass stars, the time ($t_{\rm first\; SN}$) of the first supernova (Sect.~\ref{sec:SN}), the estimated total stellar populations $\mathcal{E}[N_{\rm tot}]$ based on the X-ray detected stars and known O/B stars, the estimate of the total stellar mass (Sect.~\ref{sec:estimateclusterpop}), and the number of stars that were classified as cluster members by DBSCAN in GP25.}
    \label{tab:ClustersCarOB1Ages} 
\end{table*}

\begin{table*}
    \caption{Maximum-likelihood mean positions, distance, proper-motion velocities, and median radial velocities of each cluster.}
    \centering
\begin{tabular}{lccccccrr}
    \hline\hline
      Cluster &RA&DEC&Distance &  $\langle v_{\rm RA} \rangle$ & $\langle v_{\rm DEC} \rangle$&\multicolumn{1}{c}{$\langle v_{\rm rad} \rangle$} & $N_{\rm pm}$ & $N_{\rm rad}$   \\
      &(h:mm:ss) & (d:mm:ss) &(pc)&  \multicolumn{3}{c}{$\rm (km\,s^{-1})$} & & \\
      \hline
Tr 16&10:45:01.87 &$-59$:44:28.5&$2320\pm\;8$&$-76.20\pm0.02$&$28.89\pm0.02$&$\;  -7.17\pm 0.96$ &209&113\\ 
Tr 14&10:44:07.14& $-59$:33:13.1& $2354\pm\;7$&$-73.13\pm0.02$&$23.69\pm0.02$&$\;  -7.05\pm 0.96$ &476&90\\ 
Tr 15&10:44:40.69& $-59$:23:03.7&$2369\pm\;7$&$-69.37\pm0.02$&$23.35\pm0.02$&$\;  -3.08\pm 1.11$ &266&29 \\ 
Cr 228&10:43:57.87 &$-60$:05:45.6&$2333\pm20$&$-73.62\pm0.05$&$20.96\pm0.05$&$\; -2.07\pm 1.29$ & 42 & 7\\ 
Bo 11&10:47:04.03& $-60$:05:36.1 &$2343\pm19$&$-71.69\pm0.05$&$23.14\pm0.05$&$\;  -1.00\pm 3.95$ & 51 & 4\\ 
NGC 3324&10:37:13.69& $-58$:36:52.2 &$2388\pm16$&$-80.43\pm0.04$&$32.15\pm0.04$&$\;-8.82 \pm 2.37$  & 90 & 6\\ 
NGC 3293&10:35:47.78 & $-58$:14:32.9 &$2325\pm\;7$&$-84.12\pm0.02$&$37.36\pm0.02$&$ -14.94\pm 0.58$ &363&157\\ 
IC 2581&10:27:19.59 &$-57$:36:23.1 &$2438\pm11$&$-83.92\pm0.03$&$42.05\pm0.03$&$ \;\;\;\; 1.40\pm 2.82$ & 125 & 4\\ 
      \hline
    \end{tabular}
    \tablefoot{The uncertainties are of the mean value. The last two columns list the number of stars with available proper motions and radial velocities.}
    \label{tab:ClusterVel}
\end{table*}

\subsection{Cluster ages, massive stars, and number of supernovae
per cluster} \label{sec:clusternotes}

\paragraph{Tr 16}
is the largest cluster in Car~OB1 and consists of several subclusters \citep[see][]{CCCP-Clusters,2025A&A...695A..48G}. Including the O-type stars in the strongly obscured subgroup Tr16-SE \citep{Tr16-SE-KMOS}, it contains 33 O-type and WR stars, the luminous blue variable  $\eta$~Car, and one blue supergiant. 
We adopted a mass of $100\,M_{\odot}$ for $\eta$~Car \citep{2023MNRAS.519.5882S} and set this as our upper mass limit for the IMF extrapolation. 
The age of Tr\;16 is  $\approx 3\!-\!4$~Myr \citep{HAWKI-survey,Povich19}.

\paragraph{Tr 14}
hosts 15 O-type stars and three B supergiants. The most massive star is the spectroscopic binary HD\;93129\;A with spectral type O2~If*/WN5~+~O3.5~III and an initial mass of $100\,M_{\odot}$ \citep{2019A&A...621A..63G} for the primary star. Recent age estimations for the cluster range from $0.7^{+3.2}_{-0.6}$~Myr \citep{2025A&A...697A..39K} to $1.4^{+1.5}_{-0.7}$~Myr \citep{2024A&A...685A.100I}. We therefore adopted an age of $\simeq 1^{+3}_{-1}$~Myr, which implies that all high-mass stars should still be present.

\paragraph{Tr 15} 
contains three O-type stars (HD~93249: O9 III + B1.5 III, TYC 8626-02506-1: O9 V(n), and Tr15-20: O9 V:) and 20 stars with spectral types between B0 and B2.5. The IMF extrapolation leads to an expected initial number of $ 3\pm1$  stars with masses higher than an O9~III star. Subtracting the supergiant HDE~303310 (M3~Iab, \citet{2023A&A...671A..20B}) from this number leads to the expected number of $ 2\pm1$ supernovae that have already happened in the cluster.  
Recent age estimates for Tr~15 are in the range of $\approx 4-8$~Myr \citep{HAWKI-survey,Povich19}, and thus we assume an age of $\approx 6 \pm 2$~Myr for the cluster.

\paragraph{Cr 228}
 is located in the Southern Pillars region, and contains four O-type stars with spectral types from O7~V((f))z to O9.7 IV and one WR star (HD~93131) with spectral type WN6ha-w, for which we adopt a mass of $68\;M_\odot$ \citep{2019A&A...625A..57H} and use as an upper mass limit. 
\citet{2013A&A...558A..53K} and \citet{2011A&A...531A.141D} estimated ages of 4.8--5~Myr for the cluster, but these values are probably too high because of sample contamination by older stars from the distributed population. As the lifetime of a $68\;M_\odot$ star is 3.7~Myr \citep{2022A&A...658A.125S}, we assume Cr~228 to have an age of $\approx  4\pm1$~Myr.

\paragraph{Bo 11}
 is also located in the Southern Pillars region and hosts four O-type stars whose spectral types range from O5~Ifvar to O9.7 IV. According to \citet{2010A&A...524A..98W}, an O5~I star has a mass of $ 57\;M_\odot$. We adopt this value as the mass limit at the massive end of the cluster. 
An IR photometric age estimate for the cluster yielded an age $\approx 5$\;Myr
\citep{HAWKI-survey}. Additionally, we can approximate the maximum age of Bo~11 with the lifetime of its earliest star, HD~93632, with spectral type O5~Ifvar. As supergiants have high mass-loss rates, we can assume the star's current mass of $ 57\;M_\odot$ as the star's minimum initial mass on the main sequence for a maximum lifetime. This leads to a lifetime of $\approx 4$\;Myr \citep{2022A&A...658A.125S}, and we assume an age of $\approx 4\!-\!5$~Myr for Bo~11.

\paragraph{NGC~3324} is located north of the Carina Nebula and harbors three O-type stars, HD\,92\,206 A and B (both of spectral type O6\;V)
and CD~$-$57~3376 (spectral type O9/B0), which create the \ion{H}{II} region Gum~31.
The estimations of NGC~3324's age in the literature range from $\approx 1-2$~Myr \citep{Preibisch14, 2013A&A...558A..53K} over 6.4~Myr \citep{2021MNRAS.504..356D} and up to 11--12~Myr \citep{2020A&A...640A...1C, 2021A&A...647A..19T}. We note that these latter high values are inconsistent with the presence of O6 main-sequence stars, which have a lifetime of $ \approx 5.5$\;Myr \citep{2022A&A...658A.125S}. These inconsistent age estimates are probably caused by using optical data and a bias due to the choice of a (too) large radius for selecting cluster members: the studies with the older age estimates used selection areas of $300 - 400$\;square arcminutes, which then include a considerable number of stars from the older distributed stellar population, and are at least ten times as large as our selection area for NGC~3324 ($28.3$\;square arcminutes). Based on these considerations, we assume the age of NGC~3324 to be $ \approx 2\pm1$~Myr.

\paragraph{NGC 3293} is located at the northwestern edge of the cloud complex and contains 41 stars with spectral types between B0.5 and B3, leading to an expected initial population of $ 14\pm 3$ with earlier spectral types than B0.5. As there are three supergiants (HD 91969: B0 Iab, HD 91943: B0.7 Ib, and CP $-$57 3502: M1.5~Iab-Ib), the number of expected supernovae is $ 11\pm 3$.
In GP25, we estimated the age of NGC~3293 to be $\approx 10-15$~Myr based on \textit{Gaia} photometry. We therefore assume an age of $ \simeq 13\pm2$~Myr for NGC~3293. This implies that NGC~3293 is the oldest of the massive clusters in Car~OB1 and that its formation marks the origin of the association.

\paragraph{IC 2581} is a young cluster located at the extreme northwestern periphery of Car~OB1. It appears rather separated from the other parts of Car~OB1, but was defined as a subgroup of Car~OB1 by \citet{1995AstL...21...10M}. Its distance is somewhat larger than that of the other clusters, but still consistent with membership in Car~OB1. 
It contains eight stars between spectral types B0.5 and B2, leading to an expected number of $ 5\pm 2$ stars with masses higher than a B0.5 star. As the cluster hosts two supergiants \citep[V399 Car: A9/F0 Ia/ab and HD 90706: B2.5Ia;][]{1978AJ.....83.1081T}, the expected number of supernovae is thus $ 3\pm 2$. 
In GP25, we determined the age of the cluster to be $\approx 12$~Myr.

We note that the cluster ages are assembled from different literature sources and are sometimes averages of various
literature values. They were therefore estimated with different methods, which might lead to issues
considering the interpretation of age sequences. 
Together with the well-known, notorious general uncertainties of stellar age determinations, this implies that the values we list in Table~1 should be considered as
estimates, rather than precisely defined numbers.
However, we attempted to carefully revisit all ages,
coming to the tentative conclusion that the cluster ages used are reasonable and allow us to investigate the recent evolution and star-formation history of the region (for which the relative sequence of cluster ages is more important than the individual age values).

\subsection{Large-scale cloud structure and cluster ages}
\label{sec:agesequence}

\begin{figure*}
    \sidecaption
    \includegraphics[width = 12cm]{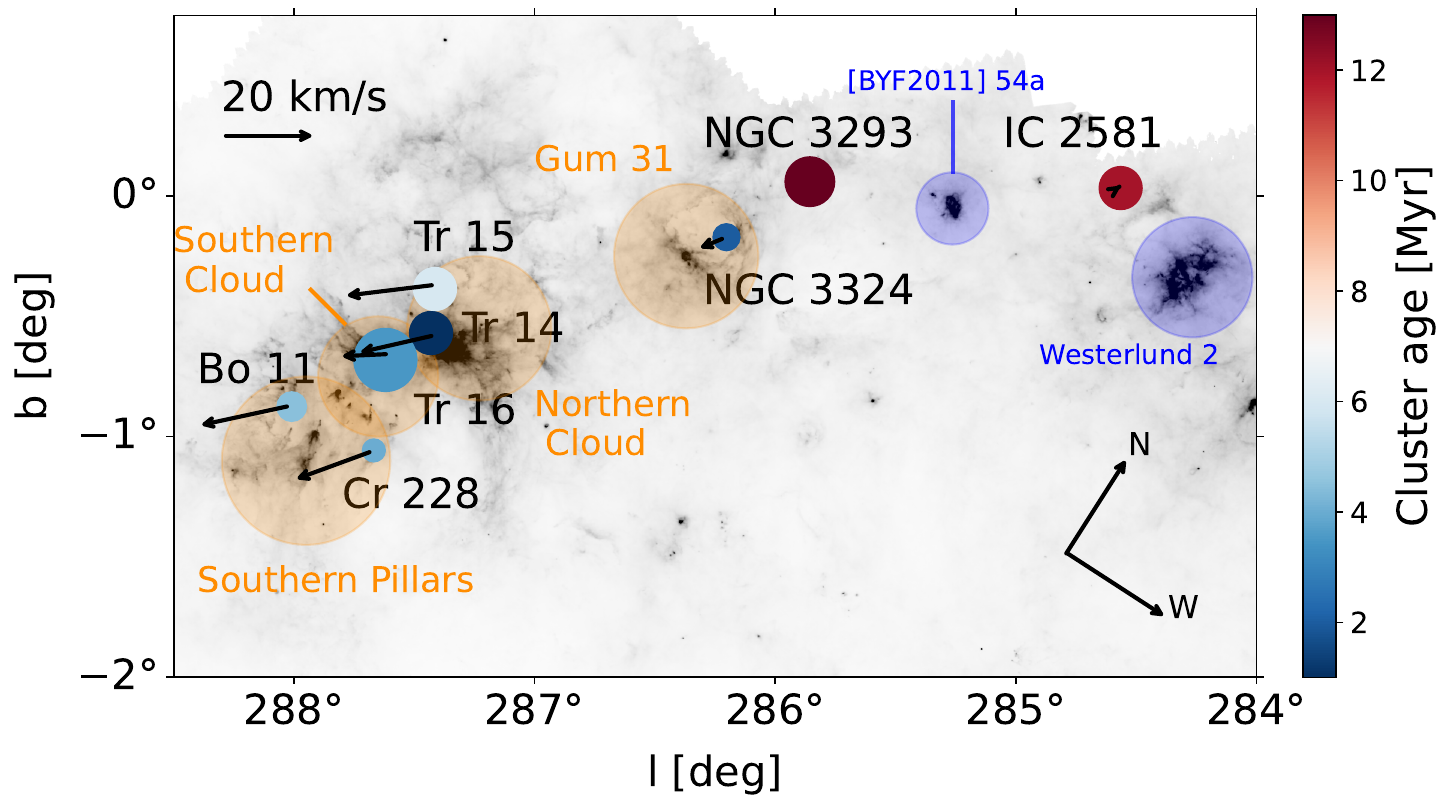}
    \caption{Observed galactic distribution of known clusters in Car OB1 (filled symbols) plotted on the \textit{Herschel} $160\;\mu \rm m$ map \citep[from \protect\url{https://www.cosmos.esa.int/web/herschel/pacs-jscanam-mosaics-hpdp}; see][]{2017ASPC..512..379G}, displayed in inverted grayscale. The sizes of the symbols are scaled by cluster radii (see Table~\ref{tab:ClustersCarOB1Ages}) and are color-coded by cluster age (see color bar).
The Carina cloud complexes are marked by orange transparent-filled circles (see orange
labels for cloud names). Moreover, unrelated background cloud complexes are marked by
blue transparent circles (BYF2011 54a, $D \approx 5.3$\;kpc; Westerlund 2, $D \approx 4 - 6$\;kpc). The direction of the arrows shows the relative direction of motion of the clusters with respect to
NGC 3293, while the length of the arrows indicates their relative speed in tangential
velocities (in this 2D plane) with respect to NGC~3293.}
    \label{fig:CarOB1Age}
\end{figure*}

Figure~\ref{fig:CarOB1Age} shows the positions and extent of the clusters as circles, color-coded according to their ages, overlaid on the \textit{Herschel} $160\;\mu \rm m$ map, which
traces the distribution of cool gas in the association. The dense clouds are largely concentrated in the central and southern part of the Carina Nebula, and around NGC~3324, whereas no dense clouds are found around or to the west of NGC~3293.
There is a large-scale trend of decreasing ages from NGC~3293 in the northwest, toward the younger clusters in the central Carina Nebula, and further on to the still ongoing star formation in the Southern Pillars.
This spatial anticorrelation between the ages of the star clusters and the presence of dense cloud material in their surroundings supports the idea that the feedback from the numerous massive stars in NGC~3293, including presumably 11 supernova explosions, has completely dispersed the original clouds in the northwestern part of the association over the last 13~Myr. 

\subsection{3D configuration of the clusters}

\begin{figure*}
\centering
   \includegraphics[width = 16.5cm,trim=0.cm 0.cm 0.2cm 1cm, clip]{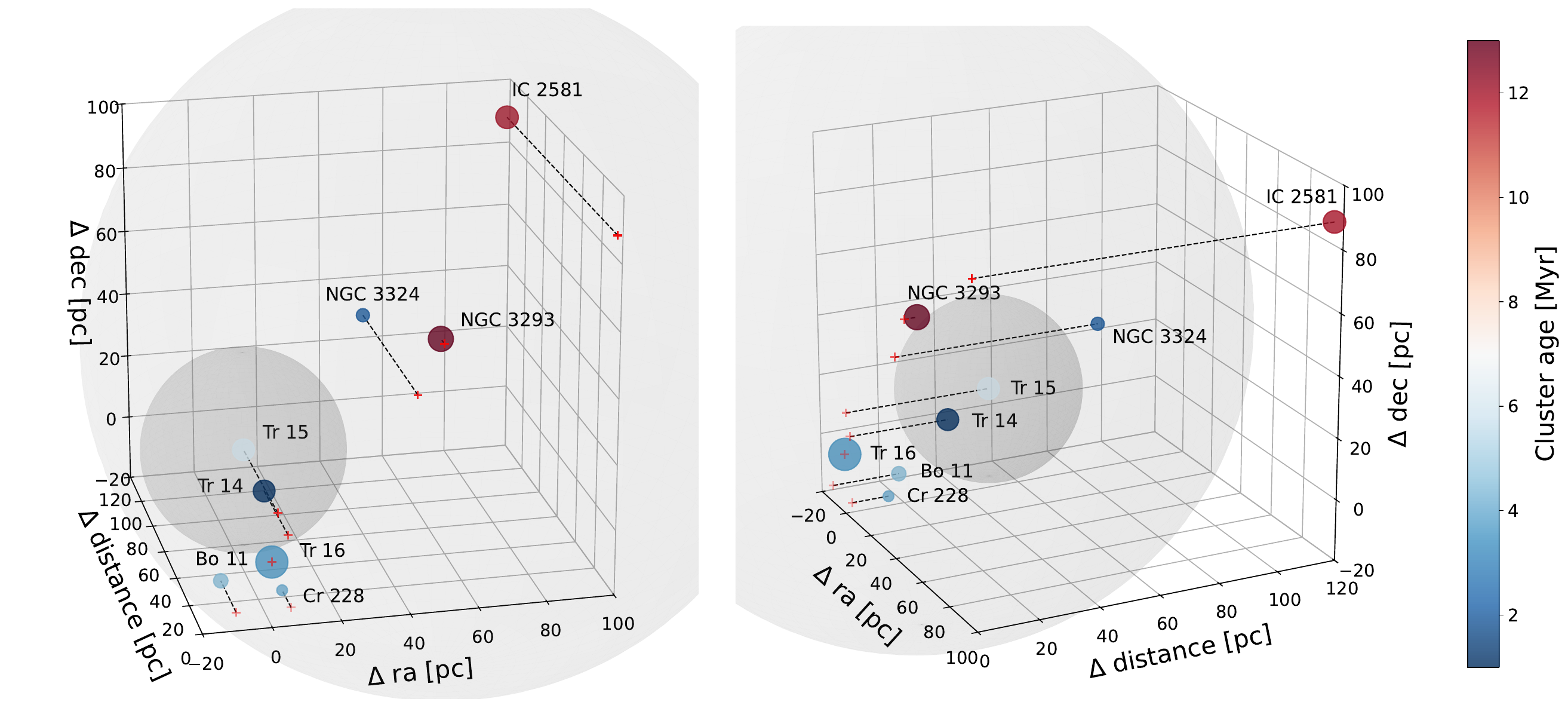} 
    \caption{Two representations of the cluster position (spheres) in 3D space, seen approximately from the front (left) and a side view (right). The red crosses show the projected cluster position on the celestial plane. The radius of the spheres is the cluster radius as listed in Table~\ref{tab:ClustersCarOB1Ages}, and the clusters are color-coded according to their ages. The gray spheres around NGC~3293 and Tr~15 represent the approximate volumes that supernova-driven bubbles originating in these clusters may have influenced over their lifetime (radii of 100~pc and 30~pc, respectively). }
    \label{fig:AgeSequence3D}
\end{figure*}

The 3D configuration of the eight clusters is illustrated in Fig.~\ref{fig:AgeSequence3D}.
The total spatial extent of the association in 3D, from Bo~11 to IC~2581, is $ \simeq 170$\,pc. 
The distances of
the three prominent clusters Tr~14, 15, and 16 in the central Carina Nebula (which appear to be in close proximity on the sky) vary\footnote{Similar differences in distances were determined by \citet{2025hsa..conf..225M}, who derived values of at $2363^{+61}_{-58}$~pc and $2354^{+61}_{-58}$~pc for Tr~14 and 15, and for Tr~16 split into east and west at $2311^{+58}_{-56}$~pc and $2305^{+64}_{-51}$~pc.} by up to $\approx 50$\;pc. Tr~14 and 15 are further away than Tr~16.
NGC~3293 is found at basically the same distance as Tr~16, while NGC~3324 and IC~2581 are 68~pc and 118~pc more distant (i.e., on the ``rear side'' of the association).

\subsection{Kinematics of the clusters}

Figure~\ref{fig:PMStars} shows the proper motion and radial velocity values of the individual stars in the different clusters. The mean values of the velocity components for each cluster are summarized in Table~\ref{tab:ClusterVel}.

Figure~\ref{fig:CarOB1Age} shows the direction of motion of the cluster in the l,b plane, using the relative tangential velocities in l,b relative to NGC~3293. It can be seen that all clusters appear to move away from NGC~3293 and that the relative 3D velocity increases with distance from NGC~3293 up to values of $\approx 25\rm\,km\,s^{-1}$, which is shown in Fig.~\ref{fig:DisvsVel} on the left. The right panel shows the velocity components, tangential and radial, for each cluster, which we calculated using equations~2--5 from \citet{2026A&A...709A.181G}.

\subsection{Relation between the clusters and the surrounding clouds}
\label{sec:relbtwclustersandclouds}

In order to investigate the physical relation between the star clusters
and the surrounding clouds, we compare the median values of the radial velocities of the stars in each cluster to the radial velocities of the molecular and atomic gas at and near the positions of the clusters, as described in detail in Section~\ref{sec:RVcompare} in the Appendix. The results of this comparison can be summarized as follows:
the (median) radial velocities of clusters Tr~14, 15, and 16, and Bo~11 are very similar to the radial velocities of the clouds around each of these clusters. This suggests that these star clusters are still kinematically related to the surrounding clouds.
For Tr~14, 15, and 16, evidence of expanding bubbles around these clusters is found in the form of gas velocity components with advancing and receding velocities with respect to the cluster velocity; the expansion velocities are around $\approx 10 \;\mathrm{km}\,\mathrm{s}^{-1} $, and provide a direct illustration of the action of massive star feedback on the surrounding gas.
 
Considering the large-scale kinematics of the molecular gas complex, variations in the mean radial velocity of
$\approx \pm 10\;\rm km\,s^{-1}$ are present, but no strong systematic velocity gradient is seen over the extent of the cloud complex \citep[on scales of $\sim 100$\;pc; see][]{2016MNRAS.456.2406R}.
However, the Northern Cloud shows a systematic radial velocity gradient on a scale of $\sim 20$\;pc, with a difference of $\approx 15\;\mathrm{km}\,\mathrm{s}^{-1}$ between the northern and southern parts of the cloud. Similar velocity differences are also seen in the clouds surrounding Gum~31.

The radial velocities of the individual clumps \citep{2005ApJ...634..476Y} and 
pillars \citep{2020MNRAS.491..178K} in the Southern Cloud and Pillars differ by up to  $ \approx 15\;\mathrm{km}\,\mathrm{s}^{-1}$.
This implies that individual clouds (clumps) can move by up to $\approx 15\;\mathrm{pc}\,\mathrm{Myr}^{-1}$, and shows that the shape and morphology of the cloud complex have probably changed considerably during the last 10\;Myr. 

\subsection{Kinematic traceback of the clusters}

\begin{figure*}
\centering
\begin{subfigure}[h]{0.33\textwidth}
     \includegraphics[width=\textwidth]{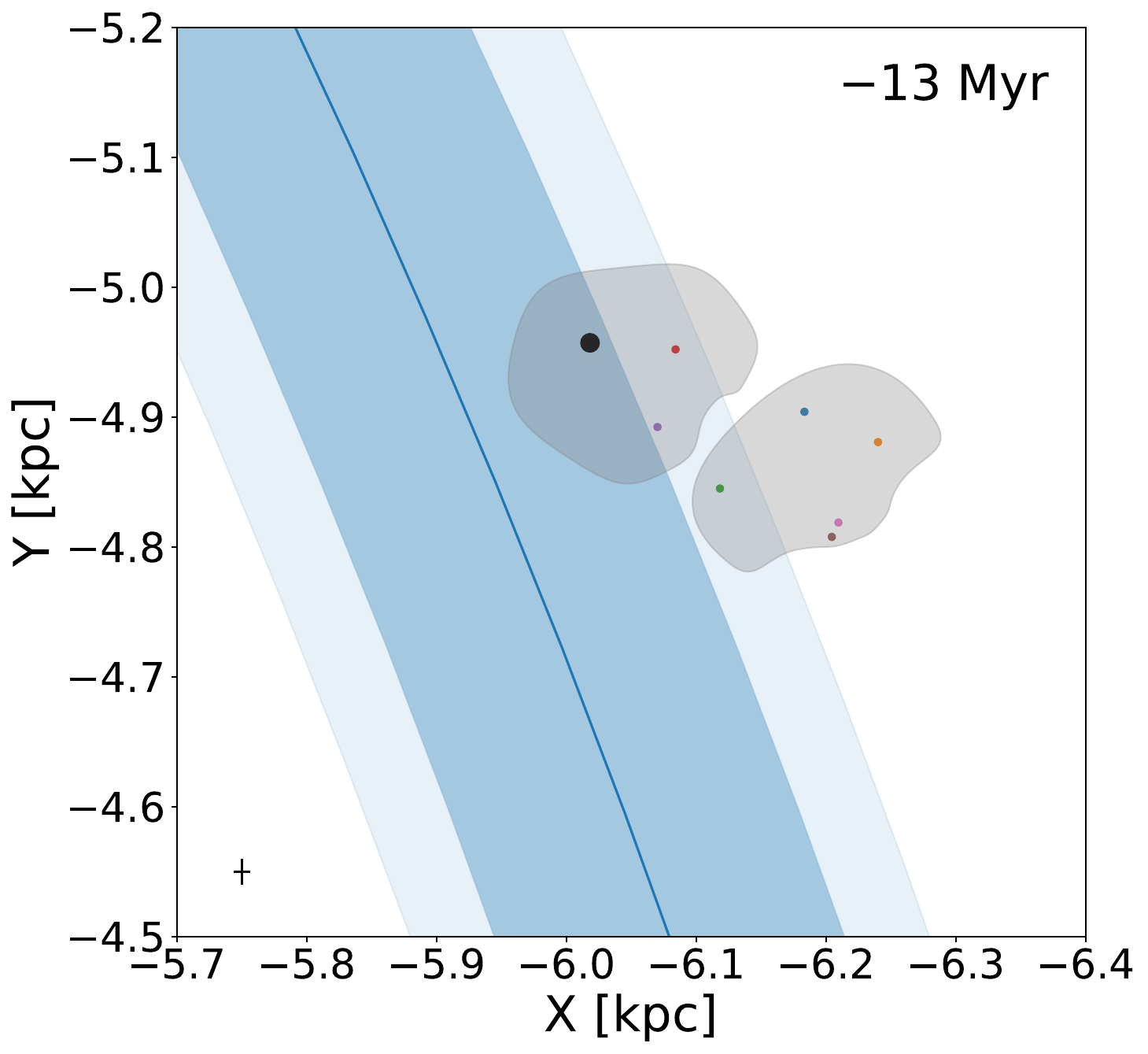}
\end{subfigure}
  \begin{subfigure}[h]{0.33\textwidth}
     \includegraphics[width=\textwidth]{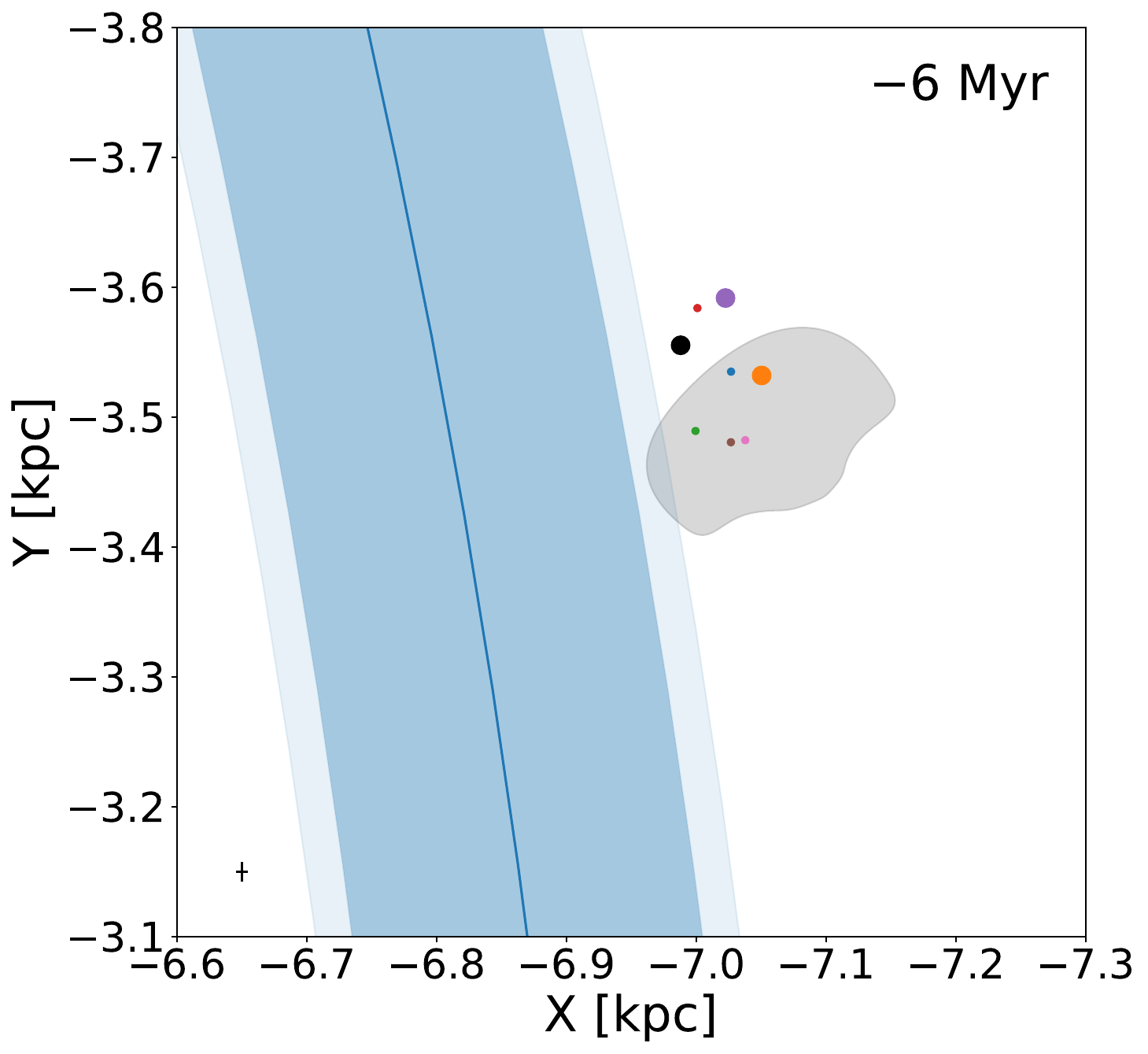}
\end{subfigure}
\begin{subfigure}[h]{0.33\textwidth}
     \includegraphics[width=\textwidth]{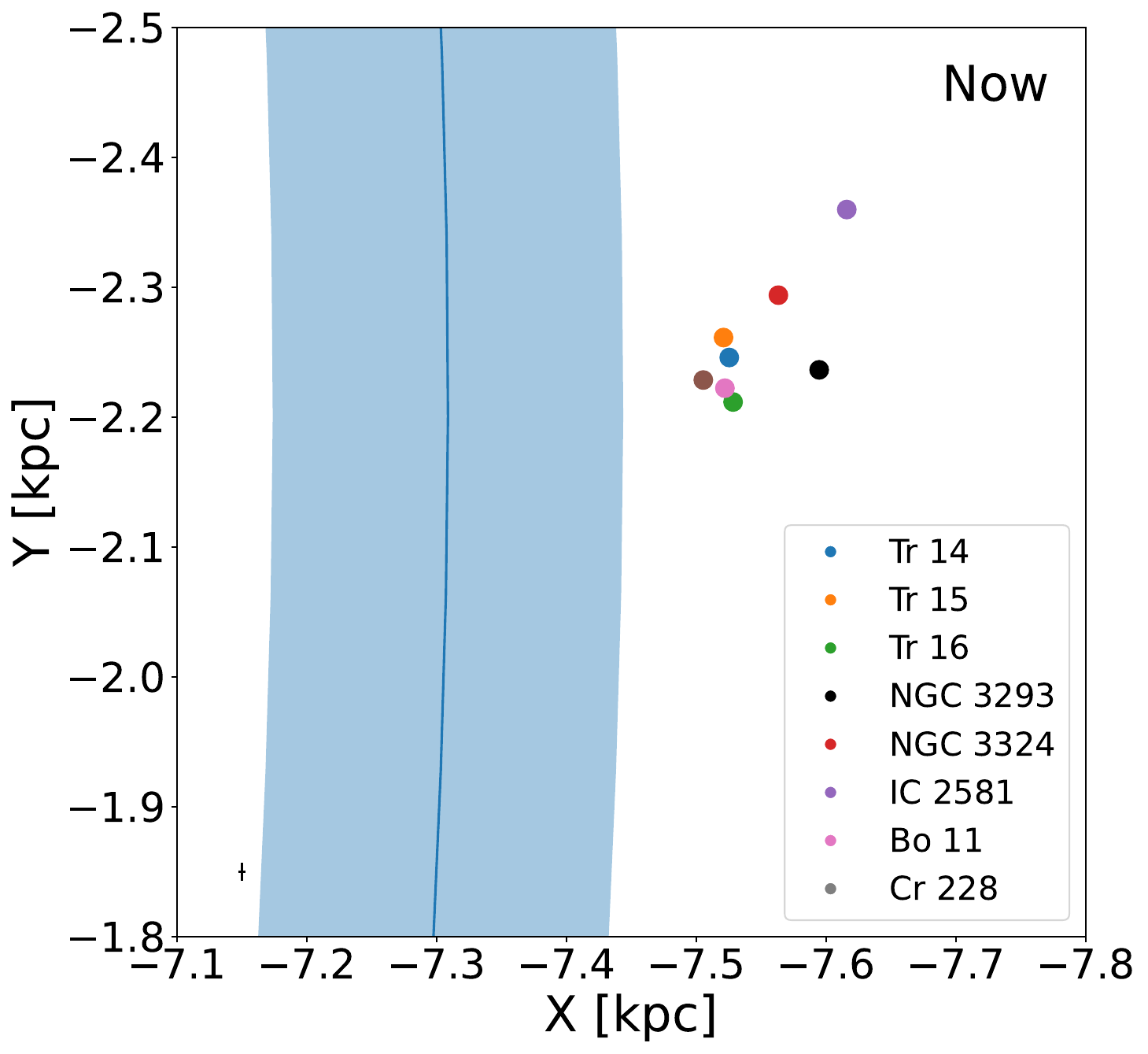}
\end{subfigure}
\caption{Galactic ``bird's eye''  view (galactic center at $(X,Y) = (0,0)$, current solar position at $(X,Y) = (-8.23, 0)$\;kpc, clockwise galactic rotation) of the traceback and current positions of 
the Car~OB1 clusters (circles), for times of 13\;Myr and 6\;Myr ago and today. If the traceback time exceeds the cluster age, the cluster is shown as a smaller and transparent cluster.
The blue line and shaded area show the 
location of the Carina spiral arm, with the lighter shaded area showing the $ 1\sigma$ uncertainty region based on the uncertainty of the pattern speed. The gray cloud-shaped areas illustrate the hypothetical positions of the two initial main cloud components out of which Car~OB1 formed (see discussion in Sect.~4).}
\label{fig:ClusterMotion_SpiralArm}
\end{figure*}

In Figure~\ref{fig:ClusterMotion_SpiralArm}, we show the positions of the Car~OB1 clusters and the Carina spiral arm in the galactic plane at three different times:
13\;Myr ago (when NGC~3293 formed), 6\;Myr ago (when Tr~15 formed), and today. 
The traceback analysis suggests that NGC~3293 was born in the spiral arm.

\begin{figure*}[h!]
\centering
\begin{subfigure}[h]{6.0cm}
   \includegraphics[width = 6.0cm,trim= 2mm 14mm 2mm 2mm, clip]{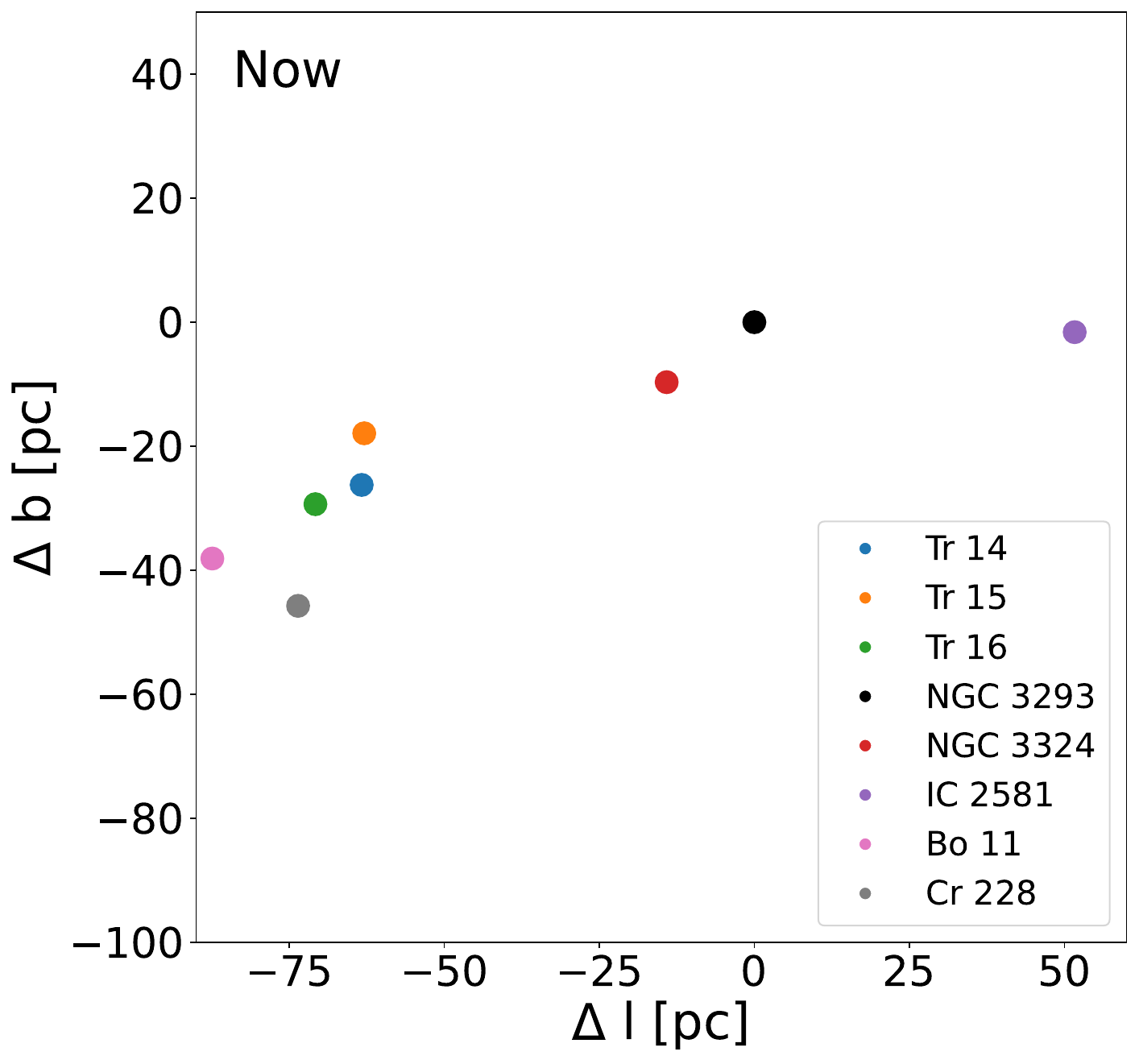}
\end{subfigure}
\begin{subfigure}[h]{5.6cm}
  \includegraphics[width=5.6cm,trim= 16mm 14mm 2mm 2mm, clip]{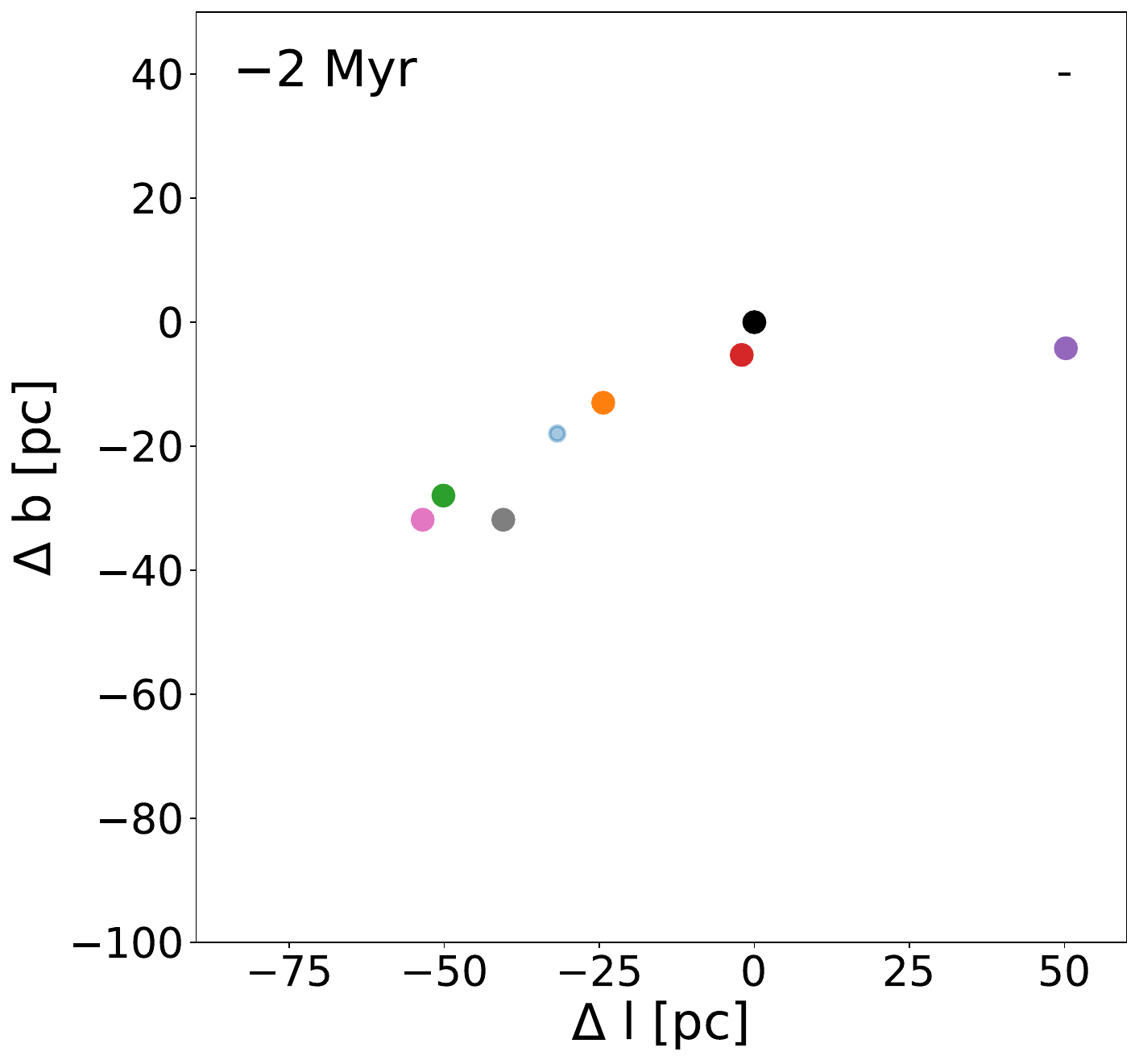}
\end{subfigure}
\begin{subfigure}[h]{5.6cm}
   \includegraphics[width=5.6cm,trim= 16mm 14mm 2mm 2mm, clip]{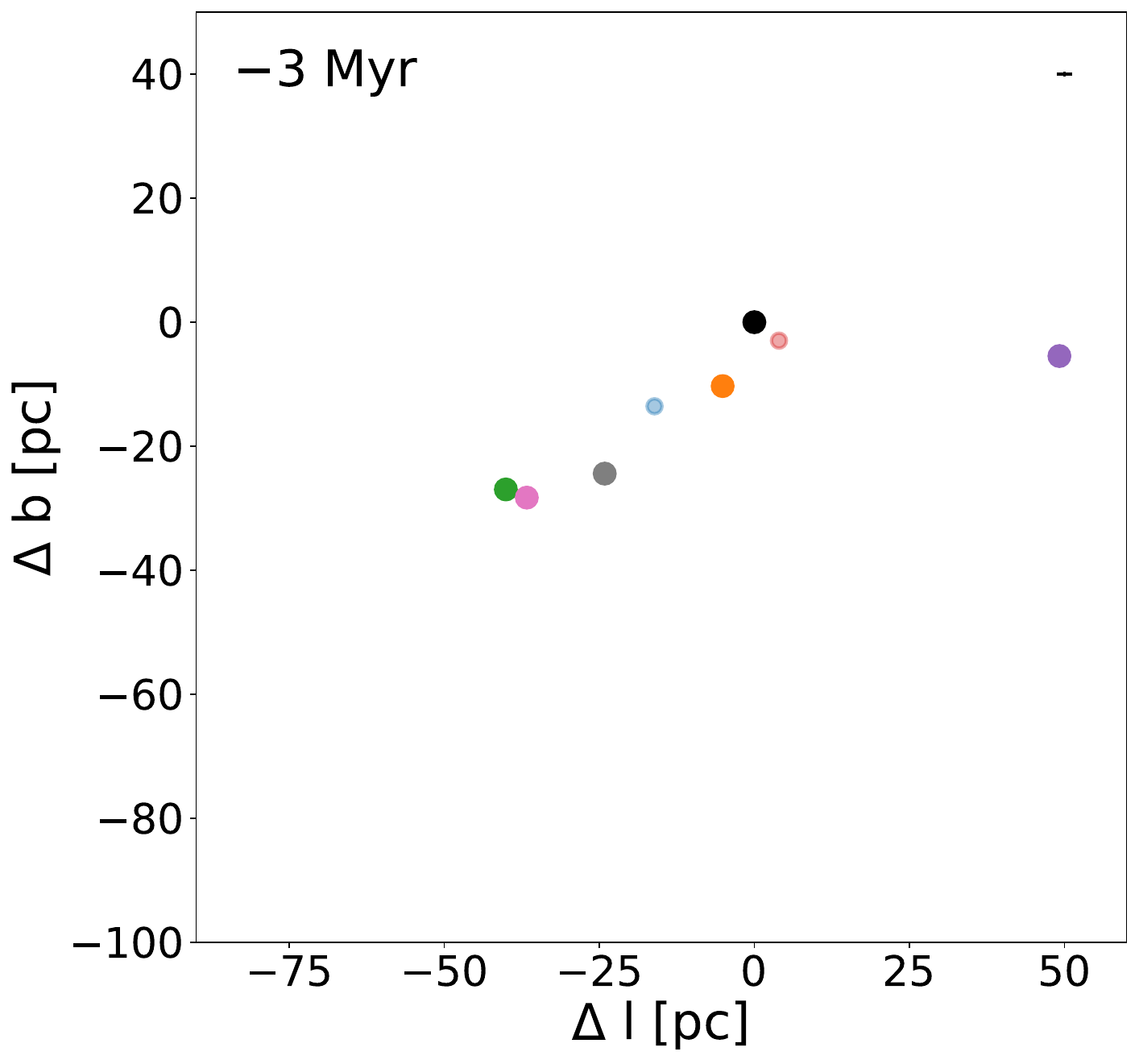}
\end{subfigure}
\begin{subfigure}[h]{6.0cm}
   \includegraphics[width = 6.0cm,trim= 2mm 2mm 2mm 2mm, clip]{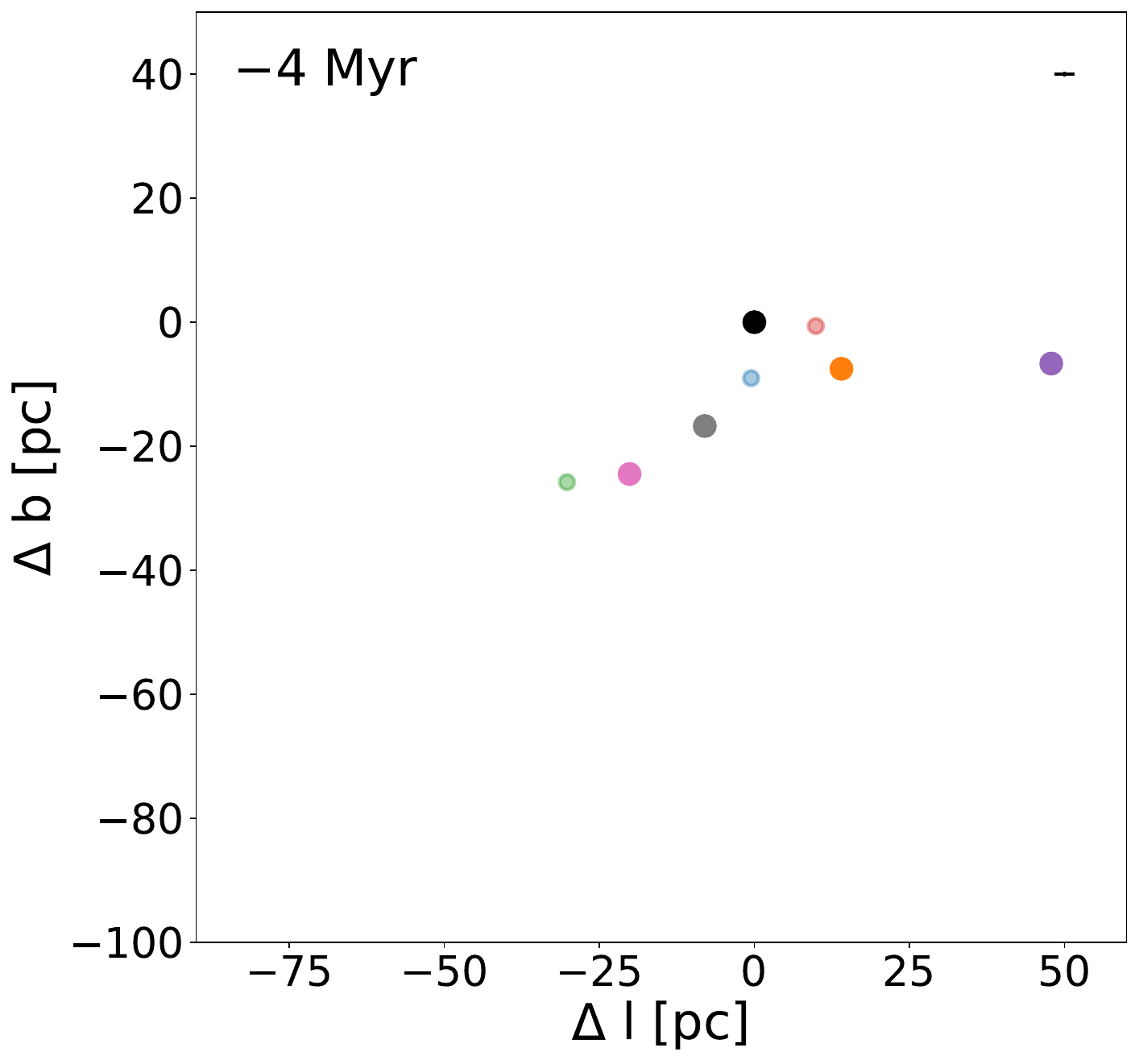}
\end{subfigure}
\begin{subfigure}[h]{5.6cm}
\includegraphics[width=5.6cm,trim= 16mm 2mm 2mm 2mm, clip]{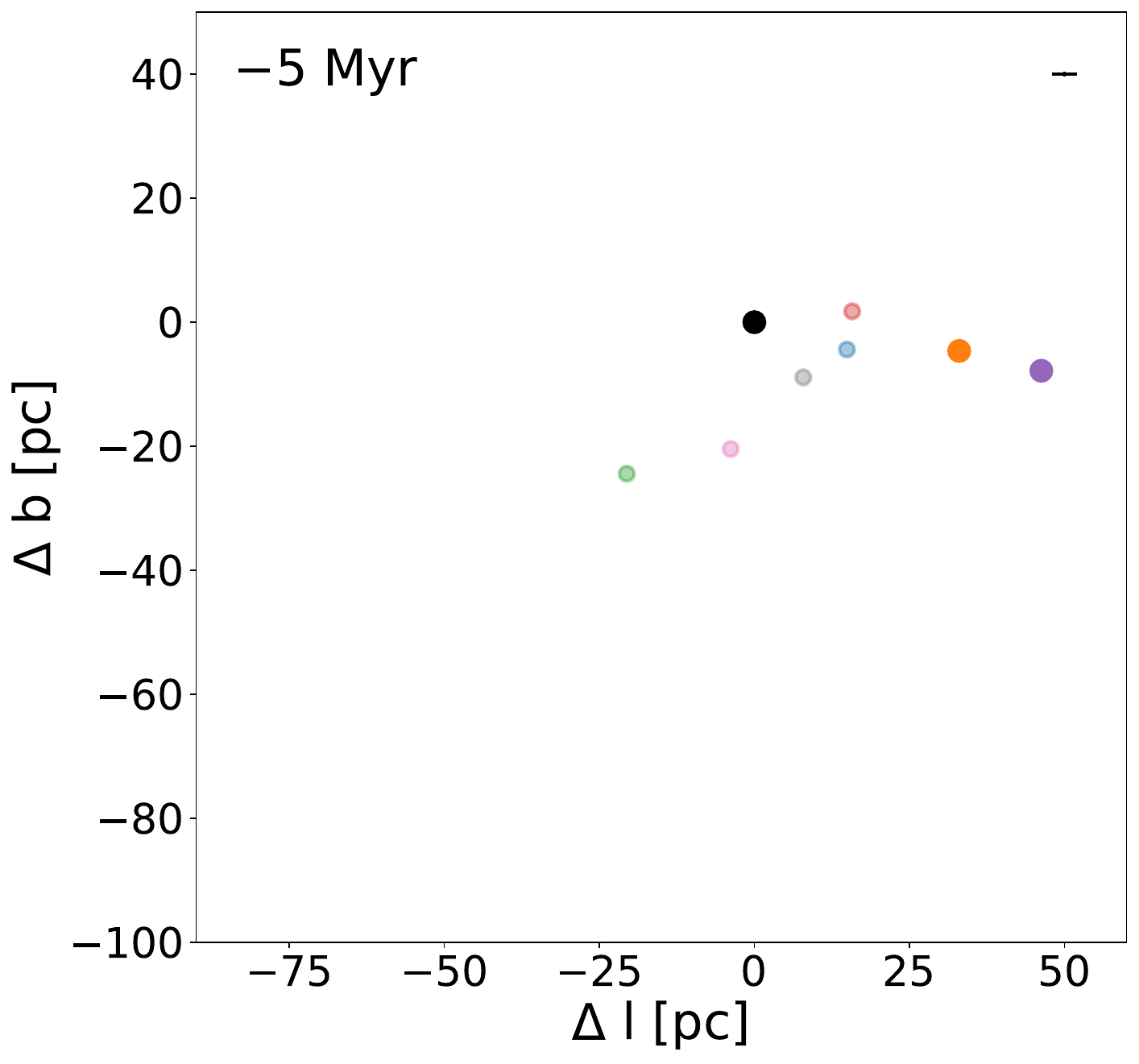}
\end{subfigure}
\begin{subfigure}[h]{5.6cm}
   \includegraphics[width=5.6cm,trim= 16mm 2mm 2mm 2mm, clip]{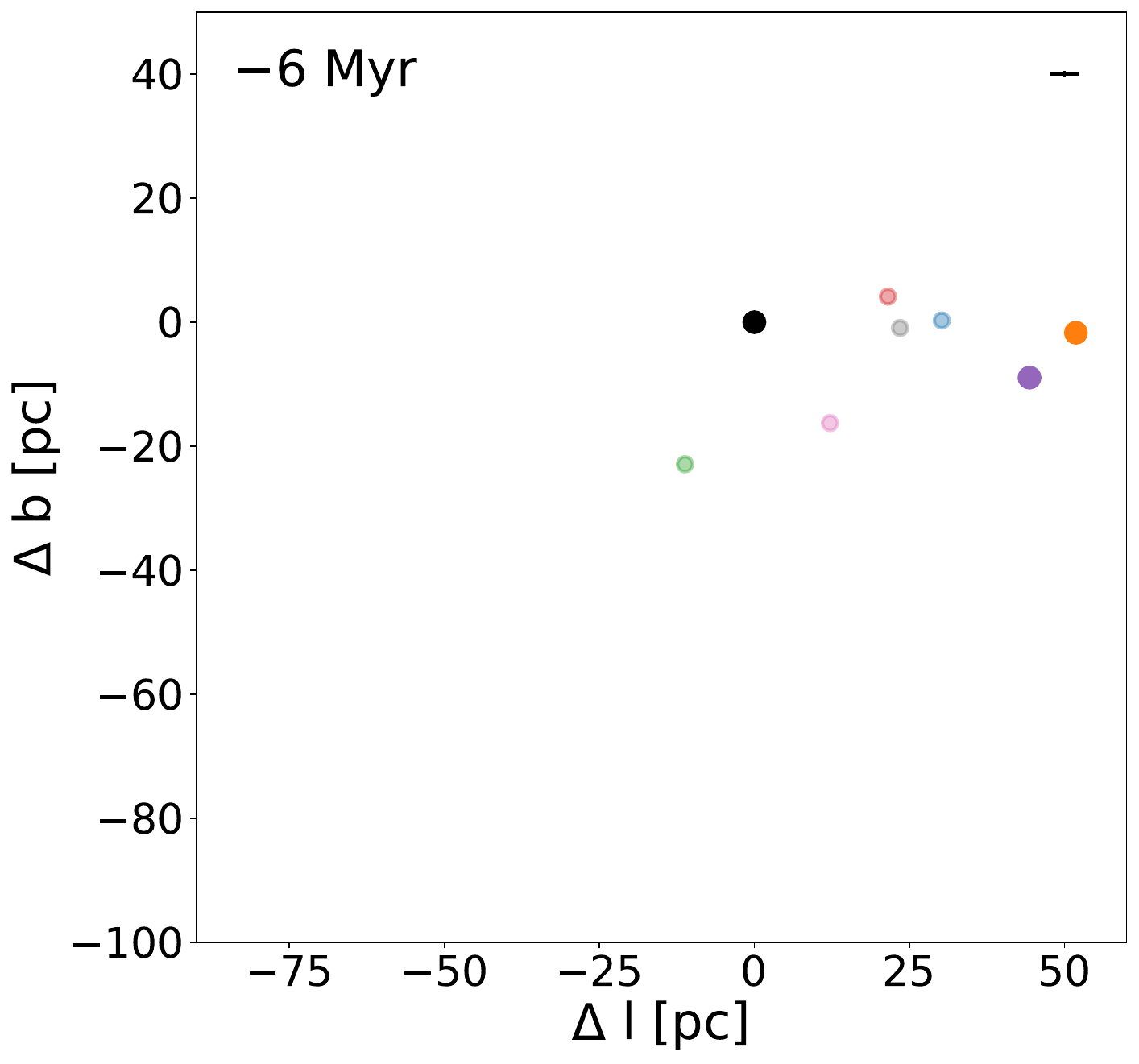}
\end{subfigure}
\caption{Traceback positions of the clusters in Car~OB1 relative to the position of NGC~3293 (fixed at (0,0)), in the plane of the sky, from today until $-6$~Myr. The cluster positions are shown by circles, which are replaced by smaller transparent circles at traceback times exceeding the cluster age. The mean relative positional uncertainties of the clusters are shown in the upper right corner.}
\label{fig:tracebackCenters}
\end{figure*}

The first panel in Fig.~\ref{fig:ClusterMotion-XY-XZ-YZ} shows the motions of the clusters in Car~OB1 relative to NGC~3293 over the last 10\;Myr in the galactic $(X,Y)$-plane, and Fig.~\ref{fig:tracebackCenters} for the last 6\;Myr in the plane of the sky. 
Today, all clusters appear to move away from NGC~3293, with velocities of about $ 8-24\;\rm km\,s^{-1}$.
Going back in time, the mean distance between the clusters gets smaller, until
the most compact configuration is found $\approx 3$--4\;Myr ago, which is illustrated in Fig.~\ref{fig:ClusterDistances}. This is consistent with our result in GP25, where we showed that the association had its smallest extent $\approx 3$--4\;Myr ago.
However, it is important to note that the pattern of expansion is not an ``explosion''-like expansion from a central point.
Rather, the traceback suggests that the clusters that today constitute the southern part of the association (or their respective progenitor clouds) moved tangentially past the cluster NGC~3293, and the minimum mean lateral distance between these clusters and NGC~3293 was reached about 3.5\;Myr ago.

Continuing the traceback to earlier times leads again to more extended configurations.
However, it has to be emphasized that the reliability of the traceback of the internal motions of the clusters within Car~OB1
gets increasingly limited if one goes back in time, since
\texttt{galpy} only uses the galaxy's gravitational potential for the calculation;
neither the influence the clusters exert on each other nor the 
gravitational interaction between the clusters and the surrounding (very massive, several $100\,000\;M_\odot$) cloud complexes is taken into account.
Therefore, while the traceback positions up to 4~Myr in the past should be reliable, the calculated positions at earlier times are increasingly uncertain. The cluster positions and velocities at each timestep are available in their entirety at the CDS; a preview of the table is shown in Table~\ref{tab:ClusterTraceback}.

\section{Discussion} 

\subsection{Triggered star formation in Car~OB1}

Today, we see two different kinds of triggered star formation currently at work
in the CNC. 
The first mechanism is radiative triggering, where the strong irradiation from the massive stars in Tr~16 and~14 compresses the clouds in the Southern Pillars \citep{Smith10b}, which leads to cloud collapse and star formation. In a highly inhomogeneous cloud complex like the Southern Pillars, this mechanism is expected to create individual stars or small stellar groups and/or clusters.

The second form of current triggering is by the expansion of an \ion{H}{II} region. The expansion collects cloud material at the edge of an \ion{H}{II} bubble and compresses it until cloud collapse and star formation set in 
\citep[``collect \& collapse model'', see][]{1977ApJ...214..725E}. This is seen at the rim of the \ion{H}{II} region Gum~31 around NGC~3324, where a dense shell has been swept up. In some parts of this rim, a very recent episode of star formation has created embedded clusters of very young stellar objects; prominent examples are the embedded cluster  G$286.38\!-\!0.26$ \citep[see][]{Ohlendorf13} at the southern rim of the \ion{H}{II} region, or the numerous embedded protostars revealed by the famous JWST image\footnote{https://webbtelescope.org/contents/news-releases/2022/news-2022-031} of the ``Cosmic Cliffs'' at the western edge of the rim \citep[see][for an analysis of the very young stellar population embedded in this rim]{2024AJ....168...63C}. 
These two mechanisms, radiative triggering and expanding \ion{H}{II} regions, are thought to produce small stellar groups or moderate-sized clusters, and are probably the main source for the widely distributed population of young stars seen throughout the whole Car~OB1 association (and in other associations).

A third way of triggering star formation is the interaction of expanding large-scale (super)bubbles, driven by stellar winds and supernova explosions in a massive cluster,  with another molecular cloud at a suitable distance \citep[e.g.,][]{ 2013A&A...550A..49K,2014A&A...566A..94K,2018A&A...619A.120K}. 
While the effect of a supernova shock wave is very destructive for the cloud in which the supernova explodes \citep[e.g.,][]{2020MNRAS.493.4700L},
there are numerous observations suggesting that supernova bubbles trigger star formation in other clouds at distances of at least $\ga 10$\;pc and up to $\la 100$\;pc. Examples have been found in nearby regions \citep{2022Natur.601..334Z}, in
other galactic OB associations \citep[e.g.,][]{2005AJ....129..393O, 2009ApJ...694.1423L,2018A&A...619A.120K},
and on galaxy scales \citep[e.g.,][]{2018MNRAS.478.3386E}. 

Supernova-driven bubbles reach typical radii of a few tens of parsecs within a few megayears \citep[e.g.,][]{2013A&A...550A..49K}. The further expansion of the bubble generally slows down with time, but each subsequent supernova increases the pressure in the bubble again. If the bubble interacts with another cloud, each pressure increase by a subsequent supernova will increase the compression of the cloud again and may drive further parts of the cloud into collapse. This is illustrated by \citet{2018A&A...619A.120K} with their ``Surround and Squash'' model, which shows how expanding superbubbles can strongly change the large-scale morphology of the clouds during the interaction. This large-scale interaction can efficiently redistribute and concentrate large amounts of gas at specific locations. If such a massive, dense cloud clump subsequently experiences a further episode of compression (since a further supernova explosion increases the pressure in the bubble), this may lead to the formation of a massive star cluster. Supernova feedback is therefore a suitable way to form rather massive star clusters \citep[e.g.,][]{2021MNRAS.504.1039R}, and a good explanation for the formation of the massive clusters in Car~OB1.
Sequential star formation for Car~OB1 was first suggested by \citet{1980AJ.....85.1193T}, who discovered an age gradient for Car~OB1 with older clusters in the northwest and theorized that the older clusters could have affected the formation of the younger clusters via feedback. 

\begin{figure*}[h!]
\centering
\begin{subfigure}[h]{6.0cm}
  \includegraphics[width=6.0cm]{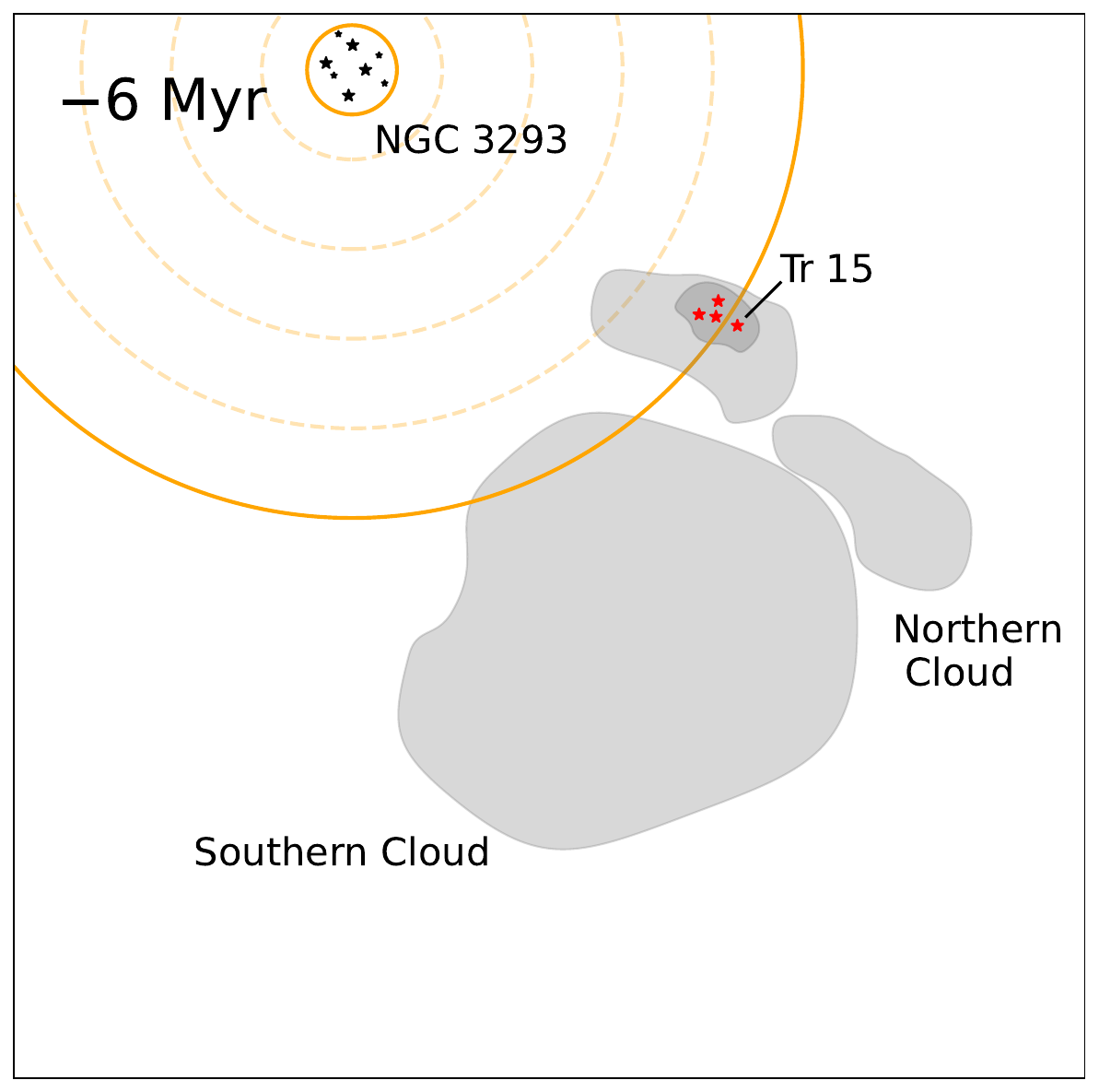}
\end{subfigure}
\begin{subfigure}[h]{6.0cm}
   \includegraphics[width=6.0cm]{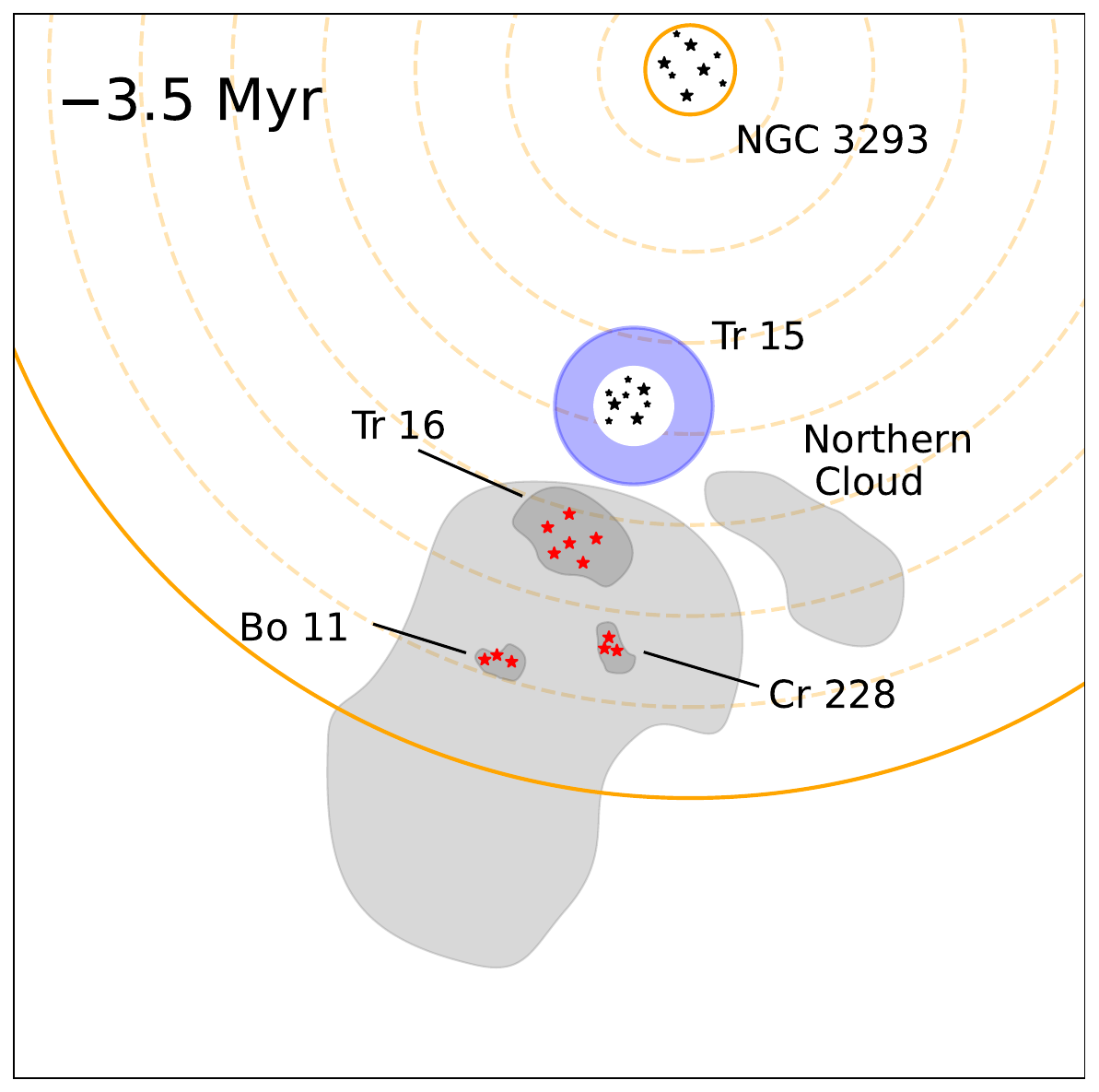}
\end{subfigure}
\begin{subfigure}[h]{6.0cm}
\includegraphics[width=6.0cm]{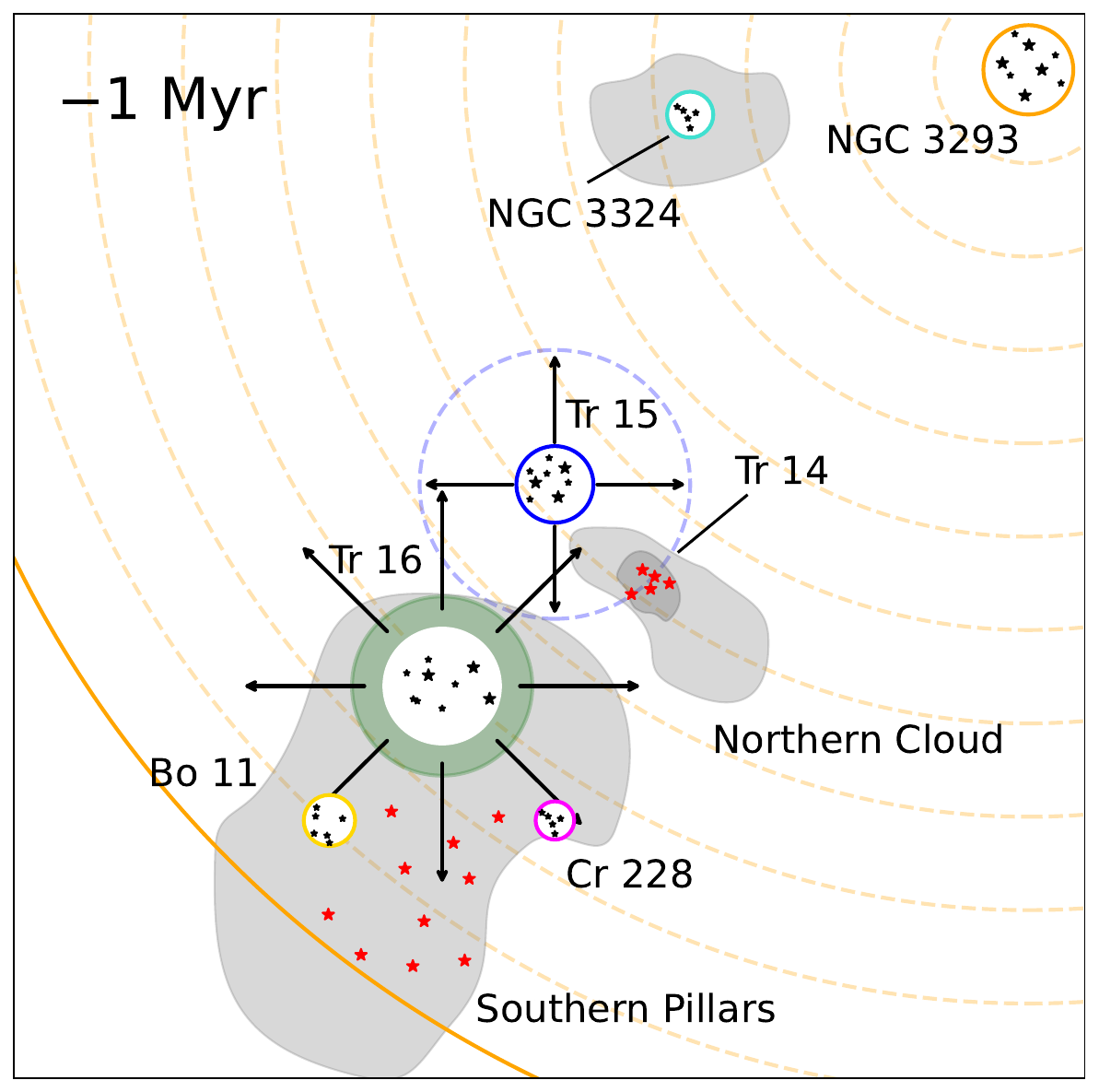}
\end{subfigure}
\caption{Illustrative sketch (not to scale) of the formation scenario for the clusters in the central part of Car~OB1, visualizing the compression and triggering effects of the expanding bubble shocks on clouds.
}
\label{fig:FormationScenario}
\end{figure*}

\subsection{A formation scenario for Car~OB1}
\label{sec:SFscenario}

Today, the total mass of the clouds observed in Car~OB1  is $\sim 10^6\;M_\odot$ \citep{SB07,Preibisch12,2016MNRAS.456.2406R}; about 10\% of this cloud mass is in the form of dense ($A_V > 3$\;mag) clouds today, while 90\% of the gas mass is in a more diffuse, low-density phase.
The total stellar population comprises about $\approx 10^5$ stars, which implies a total stellar mass of
$\approx 50\,000\;M_\odot$.

The large majority of the dense cloud mass present today is located in the southern parts of Car~OB1 (i.e., the CNC),
whereas only very little dense cloud material is seen in the northern part of  Car~OB1.
However, the morphology of the original cloud complex, before the start of star formation, must have been very different from what we see today, because the oldest clusters are located at the northern edge of the association.

Based on our results for the cluster trajectories from the traceback, cluster ages, and the estimated numbers and times of supernovae, we propose the following cluster formation scenario for Car~OB1, which is illustrated in Figure~\ref{fig:FormationScenario}.
We emphasize that the limited accuracy of the kinematic data
leads to uncertainties, and thus this scenario should not be
considered as a precise and detailed description of what actually happened, but rather
as a possible sequence of events that is consistent
with the currently available data.

\paragraph{Initial conditions}
The original cloud complex consisted of two main components: one complex out of which NGC~3293 (and perhaps IC~2581) formed, 
and another out of which Tr~14, 15, and 16, Bo~11, and Cr~228 formed. These two cloud complexes were moving past each other with relative velocities of about
\mbox{10--20\;km\;s$^{-1}$}. The cloud complex forming NGC~3293 and IC~2581
is no longer present, since the 13 supernova explosions that happened during the last 10\;Myr in this area have completely dispersed the original clouds. 
The (remnants of the) cloud complex in which Tr~14, 15, and 16, Bo~11, and Cr~228 formed now constitute the Southern Cloud, Southern Pillars, and Northern Cloud.

\paragraph{START: Formation of NGC~3293}
Star formation in Car~OB1 started $ \approx 13$\;Myr ago in the northern part of the original Car~OB1 cloud complex with the formation of NGC~3293. Our traceback analysis suggests (see Fig.~\ref{fig:ClusterMotion_SpiralArm}) that NGC~3293 formed inside the Sagittarius-Carina spiral arm when its natal cloud was compressed during its passage through the spiral arm density wave.

NGC~3293's initial population of 14 O-type stars created an expanding \ion{H}{II} region and irradiated the surrounding clouds; this stellar feedback started to create the distributed population of young stars in the northern part of Car~OB1.
After about 4~Myr (i.e., $\approx 9$~Myr ago), the most massive O-type star in NGC~3293 exploded as a supernova. This first explosion was probably soon (within $\la 1$~Myr) followed by the next supernova (11 supernovae within 8~Myr), and within a few megayears, an expanding superbubble was formed around NGC~3293.  The feedback from this series of supernovae has now completely dispersed the original cloud in which NGC~3293 had formed.
The massive clouds in the southern part of Car~OB1, which were
$\ga 20\!-\!50$\;pc away from NGC~3293 at that time, were not destroyed
by the supernovae in NGC~3293, but instead experienced a sequence of shock waves that
(i) presumably strongly changed the global structure of the original cloud complex, and
(ii) provided a series of compressions on these clouds that triggered star formation in subsequent steps.

\paragraph{STEP 2: Formation of Tr~15}
About 3~Myr after the first supernova in NGC~3293 (i.e., about 6~Myr ago), about three to four further supernovae should have exploded in NGC~3293; the compression of this expanding superbubble on a cloud (the remnants of which are now seen as the northeastern part of the Northern Cloud)
triggered the formation of Tr~15. While we do not have exact intercluster distances for this time frame, the distance between the progenitor cloud of Tr~15 and NGC~3293 was very likely smaller than today, probably between 20~pc and 40~pc. The superbubble had thus already interacted for several million years with the cloud in which Tr~15 formed, and had time to compress and accelerate this cloud.

\paragraph{STEP 3: Formation of Tr~16 (and Bo~11 \& Cr~228)}
Tr~16 formed 3--4~Myr ago, i.e~$\approx 5\!\!-\!\!6$~Myr after the first supernova in NGC~3293, and at a distance of $ 59^{+14}_{-7}$\;pc from NGC~3293. At that time, we can expect that about five to seven supernovae had already happened in NGC~3293, which sustained the superbubble and increased the pressure in the bubble. Tr~16 is located (as seen from NGC~3293) at the 
``head'', i.e., the northern edge of the Southern Cloud and Pillars cloud complex, which has a total cloud mass
of $\approx 200\,000\;M_\odot$ today.
The counter-pressure of this massive cloud complex allowed the formation of the particularly large cluster Tr~16. The somewhat smaller clusters Bo~11 and Cr~228, farther south, formed approximately simultaneously to Tr~16, presumably also by the advancing shock wave from the expanding NGC~3293 superbubble.

\paragraph{STEP 4: Formation of Tr~14}
The birth of the particularly dense and massive cluster Tr~14 about one Myr ago seems to be related to the 
 combined influence of three factors:
 \begin{enumerate}
 \item[(i)] the superbubble from NGC~3293 (at a distance of $ 56^{+6}_{-4}$~pc) compressing the cloud in which Tr~14 formed from the northwest direction. At the time when Tr~14 formed, 11 supernovae had already exploded in NGC~3293, and the bubble had already interacted with the cloud for several million years. This interaction presumably swept up a particularly dense and massive clump; 
 \item[(ii)] the first supernova that had happened in Tr~15 (at a distance of $ 14^{+14}_{-7}$~pc) about 1~Myr before the formation of Tr~14;
 \item[(iii)] the irradiation og the 35 high-mass stars in Tr~16 at a distance of $ 37^{+13}_{-13}$~pc. 
\end{enumerate}

 Assuming a typical star-formation efficiency of $\la 5\%$, the mass of the progenitor cloud of Tr~14 must have been $\ga 100\,000\;M_\odot$.
This dense cloud received pressure from the northwestern direction (from the NGC~3293 superbubble), a supernova-shock wave coming from the northeast (from Tr~15), and strong irradiation from the east (Tr~16). The very massive progenitor cloud, the remnants of which now constitute the Northern Cloud, would have provided counter-pressure for the advancing superbubble and supernova shockwaves, which increased the compressing effect of the bubble shock waves and enabled a particularly large and dense cluster to be created. 
We note that the dense cloud seen today just west of Tr~14 (which is the densest and most massive ($\sim 45\,000\;M_\odot$) cloud structure present in the entire cloud complex today)
is in the opposite direction from Tr~15, as seen from Tr~14. This may suggest that a shock wave from the direction of Tr~15 was a decisive ingredient for the formation of Tr~14 at this position.

\paragraph{STEP 5: Recent and current triggered star formation in the Southern Pillars and the Gum~31 rim}

The distributed stellar population in the Southern Pillars has a range of ages between $\la 1$\;Myr and $\approx 4$\;Myr \citep{2014ApJ...787..108G}. This can be understood by assuming that these distributed stars have been formed by the radiative triggering from Tr~16 (and Bo~11 and Cr~228) over the last 4~Myr. Since this feedback is still ongoing, it currently forms the youngest embedded populations in this area
\citep[see][]{Smith10b}.
The rim of the Gum~31 bubble constitutes the second ``hotspot'' of current star-forming activity in the association
\citep[see][]{Ohlendorf13,2024AJ....168...63C}.

\paragraph{NGC~3324:}
This $\approx 2$\;Myr-old cluster is spatially separated from the clusters in the southern part of Car~OB1, but it shares a similar kinematics: it moves in a similar direction to the southern clusters, although with a lower velocity.
The cluster appears to be related to the rather massive cloud complex
surrounding the Gum~31 \ion{H}{II} region in the southern and western direction. 
Although NGC~3324 and NGC~3293 appear to be rather close together on the sky,
their actual 3D distance is about $ 66^{+22}_{-21}$~pc.
Our traceback suggests that, at the time of its formation, NGC~3324 had a 3D separation of $ 51^{+23}_{-23}$~pc from NGC~3293; thus its formation may have been influenced by the expanding superbubble around NGC~3293.

\paragraph{IC~2581:}
This cluster appears to be an ``outlier''. It is located at the extreme northwestern periphery of Car~OB1, about 100~pc more distant than NGC~3293, and shows a distinct kinematics, moving in a completely different direction than the other clusters.
At the cluster age of $\approx 12$\;Myr, it is not surprising that no clouds related to the cluster are seen anymore. 

The traceback puts it at a distance of $ 72^{+44}_{-36}$\;pc from NGC~3293 at the time of cluster formation. This rather large separation, combined with the small ($\la 1$\;Myr) difference in ages, suggests that its formation was probably not affected by NGC~3293.

\subsection{Nature of the apparent large-scale expansion}
\label{sec:large-scale expansion}

The limited reliability of the traceback for times more than $\approx 4$~Myr in the past may raise the question of whether the currently observed apparently expanding motion of the clusters in the southern part of Car~OB1 away from NGC~3293 could alternatively be explained by the effect of stellar feedback on the cloud complex. We therefore want to check whether the series of $\approx 11$ supernova explosions in NGC~3293 might have accelerated the cloud complex radially away from NGC~3293 to the observed mean velocity of $\approx 19\,\rm km\,s^{-1}$. Figure~\ref{fig:DisvsVel} illustrates that the southern clusters are moving radially away from NGC~3293, with most exhibiting a greater radial velocity component than tangential. Additionally, the velocity increases with distance, indicating an accelerated expansion of the association. IC~2581 is not following this trend, presumably because its formation was not influenced by massive stellar feedback from NGC 3293, which is further supported by the older age of IC 2581 that is comparable to the age of NGC 3293.

Although the 3D motions of the clouds cannot be measured, the good and general agreement between the cloud and cluster radial velocities described in Sect.~\ref{sec:relbtwclustersandclouds} suggests that the clusters are still kinematically connected to their surrounding clouds. 
With a total current cloud mass of $\sim 10^6\,M_\odot$ \citep{SB07} and a mean velocity of $\approx 19\,\rm km\,s^{-1}$ (i.e., the average radial motions, relative to the chosen reference frame, which is NGC~3293), the momentum of this global expansion in Car~OB1 can be estimated as $ \approx 1.9\times 10^7\,M_\odot\,\rm km\,s^{-1}$.
Following the discussion in \citet{2022MNRAS.515..687Q}, this number can be compared with the simulation results from  \citet{2015ApJ...802...99K}, who found that ten supernovae inject a total momentum of about $2\times 10^6\,M_\odot\;\rm km\;s^{-1}$ into the surrounding clouds.
This comparison shows that the expected total momentum from the 11 supernovae in NGC~3293 is (at least) a factor of ten too low to explain the kinematics of the large-scale expansion in Car~OB1.
Therefore, the global motion of the southern clusters away from NGC~3293 cannot be explained as a global expansion due to supernova feedback (and even less so by radiative and stellar wind feedback). It appears much more likely to be related to the initial kinematics of the various parts of the original cloud complex in which Car~OB1 formed.
This supports the idea that OB associations are born in unbound cloud complexes and their expansion is a consequence of the strong original intrinsic turbulent velocity dispersion of the individual components of the cloud complex.

\subsection{Impact of feedback in other OB associations}

The impact of stellar feedback on the evolution of star-forming regions has been studied in many regions, e.g., Sco-Cen, Orion, and the Circinus Complex.
In Sco-Cen, \citet{2018A&A...619A.120K} presented a star formation scenario describing how supernova feedback has shaped and induced the formation of the subgroups of the association. This picture was expanded by \citet{2025A&A...693A.175P}, who found ``cluster chains'', approximately linear sequences of clusters in space and time, which were formed by continuous feedback from the central clusters in the region. 
In Orion, \citet{2021A&A...647A..91G} found evidence of a major feedback event, called the Orion Big Blast event, which strongly impacted the kinematics of the complex and led to its current expansion. 
In the Circinus complex, \citet{2025ApJ...985..111K} determined that one cluster had a dominant impact on the formation of several other clusters through feedback.
Overall, these studies show that feedback has a profound impact on a star-forming region and is therefore likely to be an evolutionary driver in Car~OB1 as well.

Kinematic analyses of Sco-Cen \citep{2026A&A...705A.108H, 2026A&A...709A.181G} and Orion \citep{2021A&A...647A..91G} have shown the impact of feedback also on the kinematic evolution of OB associations. While we find qualitatively similar indications for Car~OB1, i.e.,~that the expanding motion of the clusters in the association is accelerating outward (Fig.~\ref{fig:DisvsVel}), similar to what is seen in Sco-Cen \citep{2026A&A...705A.108H}, our current knowledge of the kinematics of the low-mass clusters in Car~OB1 is still too incomplete and analyses of Sco-Cen and Orion have shown that they are crucial for understanding an OB association's kinematics.

However, a detailed comparison between these star-forming regions mentioned and our analysis of Car~OB1 is difficult for several reasons. The main factor is the distance of the region: in a very nearby region like Sco-Cen ($D \approx 140$\;pc)
\textit{Gaia} not only provides much more precise astrometric data for the individual stars,
but also reaches a much better level of completeness, allowing the astrometric identification of basically the full stellar population (including the low-mass stars). In Car~OB1 (which is $\approx 17$ times further away than Sco-Cen, and where the apparent magnitude of a star with a given mass and age is therefore at least six magnitudes fainter),
the relative uncertainties of parallaxes and proper motions are much larger.

\section{Summary}

We took a detailed look at the eight most massive star clusters in Car~OB1, and proposed a star formation scenario for the association.
We have estimated the cluster member sizes based on their high-mass star population and estimated that in total $ 16\pm4$ supernovae have occurred in the Car~OB1 association during the last $\sim 9$\,Myr so far. By analyzing the ages and the relative positions and motions of the clusters within the complex, we propose a possible star formation scenario in which the cluster NGC 3293 formed first, about 13~Myr ago, in the Carina spiral arm. We propose that feedback from about 11 supernovae from NGC 3293, in combination with later SNe from Tr~15 and the irradiation from high-mass stars in the whole region, has influenced the subsequent formation of the younger massive clusters in Car OB1. This picture is further supported by orbital tracebacks, which underline the generally expanding motions of the studied, young massive clusters in the region.

However, we find that the observed global expansion of Car OB1 is probably not dominated by stellar feedback but rather by the initial intrinsic kinematics of the original clouds. This follows from the momentum analysis, which suggests that more energy would have been needed to explain the currently observed relative motions of the young clusters.  Therefore, our results add support to the notion that OB associations form in unbound cloud complexes and that their large-scale expansion is (at least partially) due to the initial dynamics of the clouds \citep[see, e.g.,][]{2020MNRAS.495..663W}.

In the future, more precise data will become available (e.g., future \textit{Gaia} data release, additional radial velocity surveys like 4MOST), and therefore more complete membership lists will improve the study of the clustered nature of Carina, in combination with more sophisticated clustering algorithms. This will allow for more detailed analyses of massive and also low-mass clusters and distributed stars, their ages, distances, and 3D space motions. In particular, including low-mass and distributed stellar populations in the Car~OB1 association will be crucial to understanding the complex’s full evolution and the impact of feedback on these populations. Additionally, it will be possible to study the substructure of individual massive clusters, which will allow us to better understand their formation and the impact of feedback from massive stars.

\section*{Data availability}
Tables~\ref{tab:CarOB1HighMass} and \ref{tab:ClusterTraceback} are only fully available in electronic form at the CDS via anonymous ftp to \url{cdsarc.u-strasbg.fr (130.79.128.5)} or via \url{http://cdsweb.u-strasbg.fr/cgi-bin/qcat?J/A+A/}.

\begin{acknowledgements}
This research was supported by the Excellence Cluster ORIGINS which is funded by the Deutsche Forschungsgemeinschaft (DFG, German Research Foundation) under Germany’s Excellence Strategy - EXC-2094 - 390783311.
This work has made use of data from the European Space Agency (ESA) mission
{\it Gaia} (\url{https://www.cosmos.esa.int/gaia}), processed by the {\it Gaia}
Data Processing and Analysis Consortium (DPAC,
\url{https://www.cosmos.esa.int/web/gaia/dpac/consortium}). Funding for the DPAC
has been provided by national institutions, in particular the institutions
participating in the {\it Gaia} Multilateral Agreement. 
This research has made use of the open-source Python packages \texttt{Astropy} \citep{2022ApJ...935..167A}, \texttt{NumPy} \citep{harris2020array}, \texttt{scipy} \citep{2020SciPy-NMeth}, \texttt{scikit-learn} \citep{scikit-learn}, \texttt{pandas} \citep{mckinney-proc-scipy-2010}, and \texttt{Matplotlib} \citep{2007CSE.....9...90H}.
\end{acknowledgements}

\bibliographystyle{aa}
\bibliography{ref}

\begin{appendix}

\section{Relation between the clusters and the surrounding clouds \label{sec:RVcompare}}

In order to investigate the physical relation between the star clusters
and the surrounding clouds, we compare the radial velocities.
For this, one has to translate the radial velocities
reported in the literature for molecular emission lines from clouds,
which are usually given relative to the ``local standard of rest'' (LSR), to the barycentric velocity system, in which the
stellar radial velocities are given.
In the direction of the Car~OB1 clusters, the typical difference between the
LSR and barycentric radial velocity \citep[see][for details of the computation]{2009ApJ...700..137R} is $\simeq 11.6\;\mathrm{km}\,\mathrm{s}^{-1}$with individual values at the positions of the discussed clusters ranging from $11.48\;\mathrm{km}\,\mathrm{s}^{-1}$ for Bo~11, over $11.56\;\mathrm{km}\,\mathrm{s}^{-1}$ for Tr~16, to $11.89\;\mathrm{km}\,\mathrm{s}^{-1}$ for NGC~3293.

In the following, we use barycentric radial velocities and consider the
large-scale maps of molecular (CO) gas from  \citet[][R16]{2016MNRAS.456.2406R},  atomic (\ion{H}{I}) gas from \citet[][R17, R21]{2017MNRAS.472.1685R,2021ApJ...909...93R}, the velocities for the molecular clumps and pillars
reported in \citet[][Y05]{2005ApJ...634..476Y} and \citet[][K20]{2020MNRAS.491..178K}, and build on the results already reported by
\citet{2018MNRAS.477.2068K} for the central Carina Nebula. We find that the velocity differences in the pillars appear more or less random, as no strong systematic gradient is seen in the South Pillars.

\paragraph{Tr~14:}

The median radial velocity of the stars in Tr~14 of
$\langle v_{\rm rad} \rangle = {-7.1}\;\mathrm{km}\,\mathrm{s}^{-1}$
agrees well with the main peak radial velocity of the molecular gas in the Northern Cloud
$v_{\rm rad} \simeq {-8.2} \;\mathrm{km}\,\mathrm{s}^{-1}$ (R16),
and also with the range of radial velocities of the CO clumps
in the Northern Cloud of $[{-8.3}\; \dots \; {-6.0}]\;\mathrm{km}\,\mathrm{s}^{-1}$ (Y05).

The HI gas around Tr~14 shows radial velocity components at
${-14.4} \;\mathrm{km}\,\mathrm{s}^{-1}$  and ${+8.6} \;\mathrm{km}\,\mathrm{s}^{-1}$ (R17), tracing an expanding bubble around Tr~14,
probably driven by the feedback from the massive stars. This gas is flowing away from the cluster with relative velocities of $\approx 7 - 15\;\mathrm{km}\,\mathrm{s}^{-1}$. The expanding bubble is also seen in the radial velocities of ionized gas ($\approx {-22} \;\mathrm{km}\,\mathrm{s}^{-1}$  and $\approx {+10} \;\mathrm{km}\,\mathrm{s}^{-1}$)
\citep[see Fig.~11 and 12 in][]{2018MNRAS.477.2068K}. 

\paragraph{Tr~16:}
The mean velocities of the molecular gas in the Southern Cloud ($-13 \;\mathrm{km}\,\mathrm{s}^{-1}$) 
and in the Southern Pillars  ($-10 \;\mathrm{km}\,\mathrm{s}^{-1}$)  (R16)
 are slightly more negative than the median star cluster velocity of $v_{\rm rad} = -7.18\;\mathrm{km}\,\mathrm{s}^{-1}$.

The individual CO clumps (Y05) and pillars (K20) in the cloud complex show ranges of radial velocities of  $[{-{14.2}}\, \dots\,  {-0.5}]\;\mathrm{km}\,\mathrm{s}^{-1}$  and $[{-14}\, \dots\, {+6.6}]\;\mathrm{km}\,\mathrm{s}^{-1}$,

The atomic (HI) gas around Tr~16 shows components at 
${-18.4} \;\mathrm{km}\,\mathrm{s}^{-1}$ and 
${+6.6} \;\mathrm{km}\,\mathrm{s}^{-1}$ (R21), indicating an expanding shell around Tr~16 with an outward velocity of $\approx 10\;\mathrm{km}\,\mathrm{s}^{-1}$, which is 
probably driven by the feedback from the massive stars.
This expanding shell is also seen in ionized gas at velocities of
$\approx {-20} \;\mathrm{km}\,\mathrm{s}^{-1}$ and 
$\approx {+10} \;\mathrm{km}\,\mathrm{s}^{-1}$
\citep[see][]{2018MNRAS.477.2068K}.

\paragraph{Bo~11:}
The (poorly constrained) cluster radial velocity of $ {-4.25}\;\mathrm{km}\,\mathrm{s}^{-1}$ agrees approximately  with the mean velocity of the surrounding
molecular gas of $ {\approx {-10} \dots {-5}}\;\mathrm{km}\,\mathrm{s}^{-1}$ (R16).

\paragraph{Cr~228:}
The CO map of R16 does not show a significant detection of CO emission at the position of the cluster.
The (poorly constrained) cluster radial velocity of $ {-2}\;\mathrm{km}\,\mathrm{s}^{-1}$ is higher than the typcial mean velocity of the surrounding
molecular gas of $ \approx {-14}\;\mathrm{km}\,\mathrm{s}^{-1}$ (R16).

\paragraph{Tr~15:}
The molecular gas near Tr~15 shows peaks at radial velocities of  $ {-8.4}, {-3.4}$, and ${+3.6}\;\mathrm{km}\,\mathrm{s}^{-1}$ (R16, Fig.~4). Given the 
 median cluster radial velocity of  $\langle v_{\rm rad}\rangle  = -2.72\;\mathrm{km}\,\mathrm{s}^{-1}$,
this suggest an expanding shell around the cluster, \citep[as discussed by][]{2018MNRAS.477.2068K}, 
where gas is blown away from Tr~15 with an outward velocity of $\approx 5-6\;\mathrm{km}\,\mathrm{s}^{-1}$.

\paragraph{NGC~3324:}
The (poorly constrained) cluster radial velocity of $\langle v_{\rm rad}\rangle \approx  {-12.8}\;\mathrm{km}\,\mathrm{s}^{-1}$ is similar to the
general molecular gas velocity in the Gum~31 cloud of
$\approx {-10.4}\;\mathrm{km}\,\mathrm{s}^{-1}$ (R16), as well as the velocity
range of $[{-12.4}\, \dots\, {-8.4}]\;\mathrm{km}\,\mathrm{s}^{-1}$ for the CO clumps in this area (Y05).
The atomic (HI) gas shows components at 
${-8.4} \;\mathrm{km}\,\mathrm{s}^{-1}$ and
${-2.4} \;\mathrm{km}\,\mathrm{s}^{-1}$ (R21), suggesting that the gas 
moves away from us relative to the cluster. 

However, given the large uncertainty of the cluster radial velocity and the fact that the radial velocities in the Gum~31 region show a very complex pattern with strong variations (R16), this result should be treated with caution.

\paragraph{NGC~3293:}
The CO maps show a very complicated velocity distribution at the position of NGC~3293. The CO emission at that position is, however, probably strongly dominated by gas from the extended Gum~31 cloud as well as clouds in the galactic background.
Considering that about 11 supernova explosions have 
happened in NGC~3293 over the last ten Myr, their impact should have
dispersed the natal cloud of NGC~3293.

\section{Supplemental Tables and Plots}
\FloatBarrier

The sources for the collection of the radial velocities used in this study are listed in Table~\ref{tab:RVAppendix}.

\begin{table}[h!]
 \caption{References for the radial velocities used in this study, how many stars had radial velocities in each survey, and how many satisfied the applied quality cuts. The radial velocities are prioritized from the surveys in the order given in the table.}
    \centering   
     \resizebox{\hsize}{!}{\begin{tabular}{llcc}
    \hline\hline
      Survey & Reference & Counterparts & Used Counterparts  \\
         \hline
      GES DR5   & \citet{2023yCat..36760129H} & 394 & 370\\
      GALAH DR4 & \citet{2025PASA...42...51B} & 23 & 6\\
      Gaia DR3 & \citet{Katz2023} & 52 & 5\\
      & \citet{2018MNRAS.477.2068K} & 12 & 8 \\
      & \citet{2018AJ....155..190H} & 51 & 6\\
      Extended Hipparcos & \citet{2012AstL...38..331A} & 6 & 1\\
      & \citet{1969MNRAS.146..101L} & 4 & 2 \\
    \end{tabular}}
    \label{tab:RVAppendix}
\end{table}

Table~\ref{tab:SNtime} contains the estimations of the mass of the most massive star in each cluster and when it exploded as a supernova. Figure~\ref{fig:PMStars} shows the 3D motion of the clustered stars in the association, while the second and third panels of Fig.~\ref{fig:ClusterMotion-XY-XZ-YZ} illustrate the relative motion of the clusters compared to NGC~3923 in the X-Z and Y-Z plane. Figure~\ref{fig:DisvsVel} (see Sect.~\ref{sec:large-scale expansion}) shows on the left panel the relative velocity of each cluster with respect to NGC~3293 and their distance to NGC~3293. The right panel contains the tangential and radial velocities of each cluster, which were calculated using equations~2--5 in \citet{2026A&A...709A.181G}, and shows that the radial velocity component exceeds the tangential one in most clusters. Figure~\ref{fig:ClusterDistances} shows the inter-cluster distances of the clusters with NGC~3293 over the last 10 Myr. It illustrates that the distances of the southern clusters were at minima between 3 and 4~Myr ago.  

\begin{table*}[]
    \caption{Estimates for the mass and lifetime of the (initially) most massive star in each cluster, and the time of the first supernova explosion.}
    \centering
    \begin{tabular}{cccccccr}
    \hline\hline
     & & &\multicolumn{2}{c}{$M_{cl}-M_{*,\rm max}$-relation} & \multicolumn{2}{c}{IMF} & time of first SN\\
       Cluster & $\mathcal{E}[N_{\rm SN}]$ & age & $M_{*,\rm max}$  & $\tau$  & $M_{*,\rm max}$  & $\tau$ &  \\
        & & (Myr) & ($M_{\odot}$)  & (Myr)  & ($M_{\odot}$)  & (Myr) &  (Myr from today)\\
       \hline
        NGC 3293 & 11& $\approx 13$& 43 & $ \approx4.8$ & 64 & $ \approx 3.8$ & $\approx -9$\\
        IC 2581 &3& $\approx 12$& 29 & $ \approx 6.4$ & 40 & $ \approx 5.0$ & $ \approx -6$\\
        Tr 15 &2& $\approx 6$& 39 & $ \approx 5.1$ & 57 & $ \approx 4.0$ & $\approx -2$\\
         \hline
    \end{tabular}   
    \label{tab:SNtime}
\end{table*}

\FloatBarrier

\begin{figure*}[h!]
\addtocounter{figure}{-1}
\sidecaption
\begin{subfigure}[h]{0.33\textwidth}
  \includegraphics[width=\textwidth]{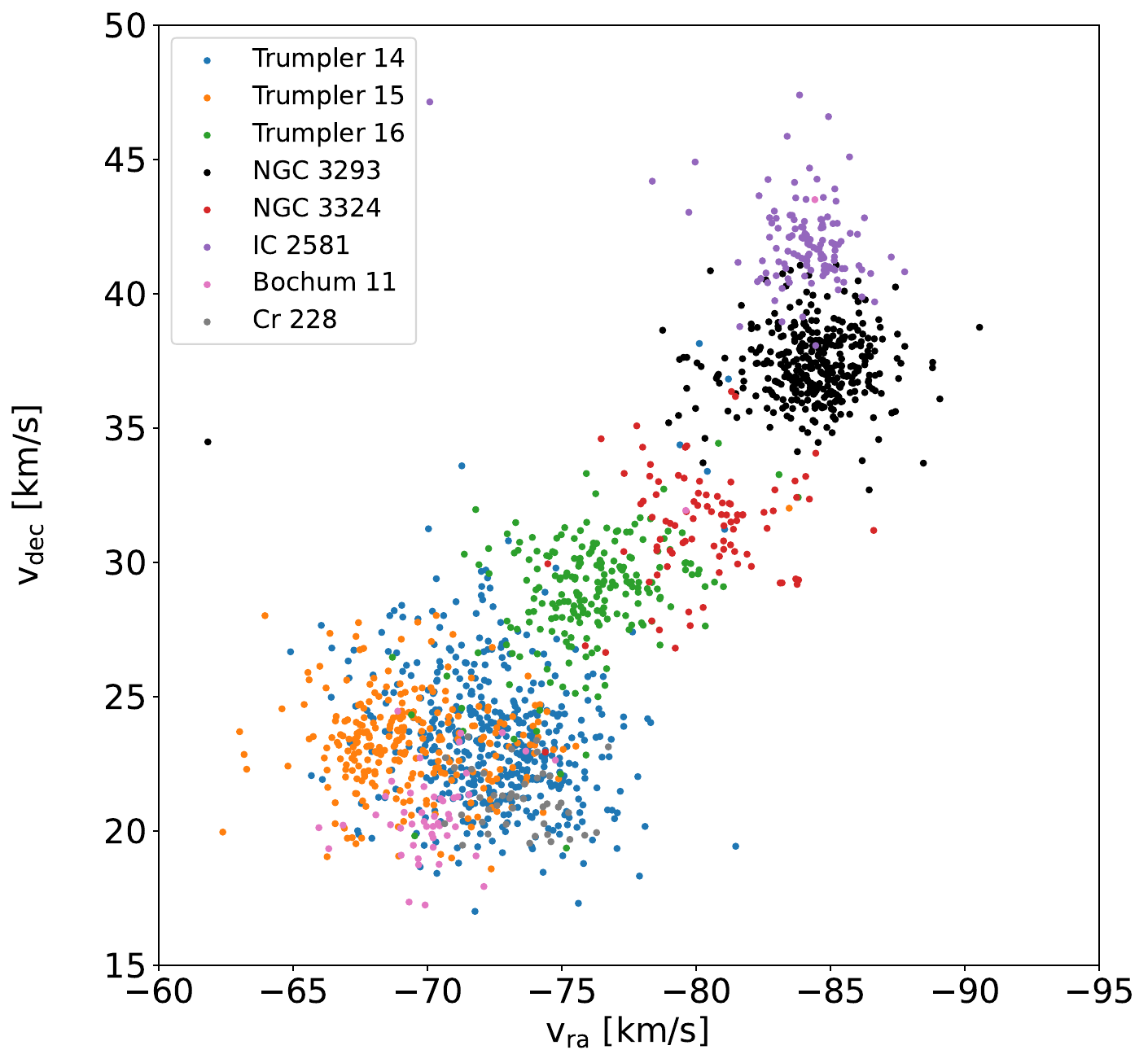}
\end{subfigure}
\begin{subfigure}[h]{0.33\textwidth}
   \includegraphics[width=\textwidth]{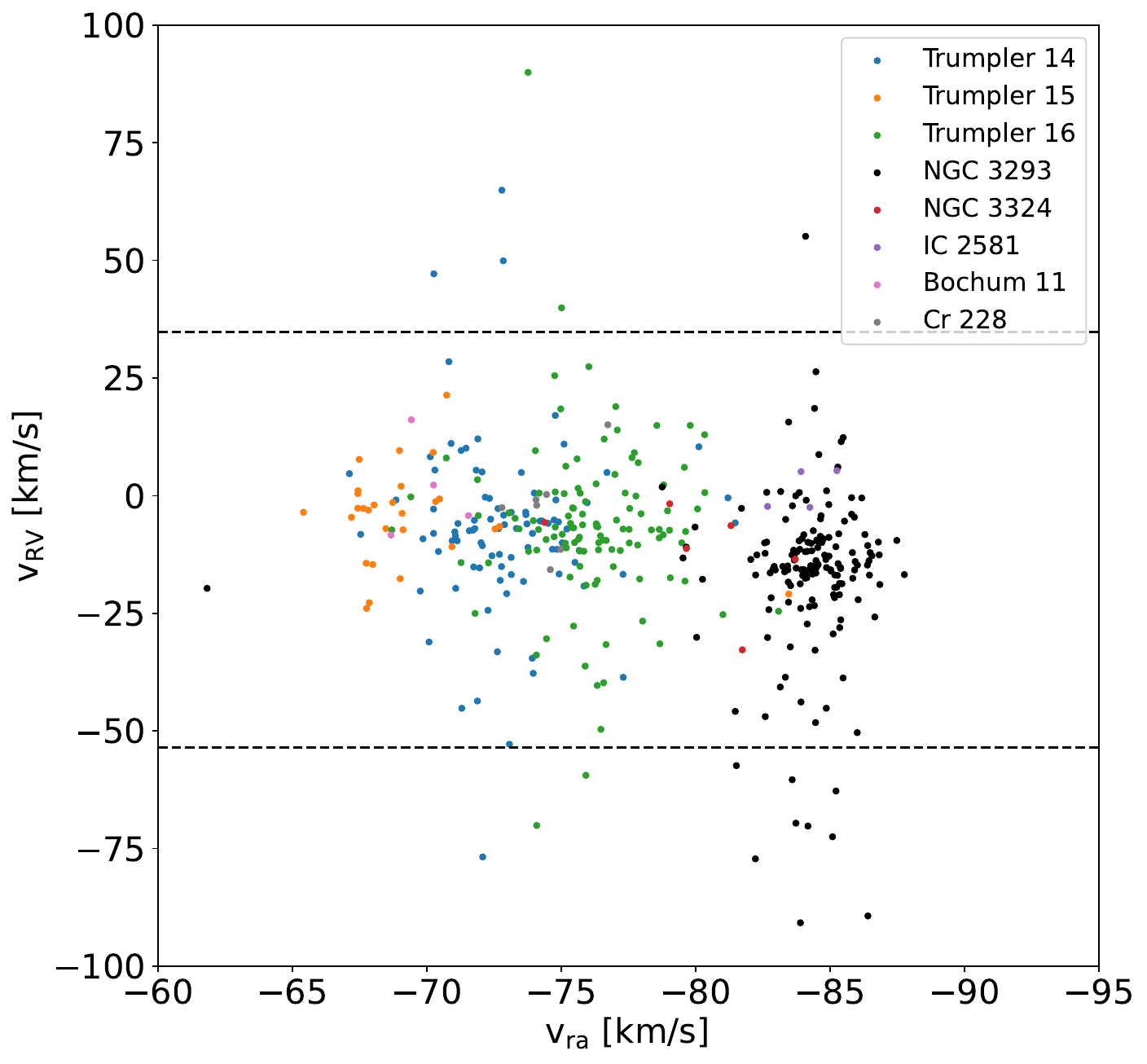}
\end{subfigure}
\caption{Left panel: Tangential velocities of the cluster member stars, colored by cluster membership as given in the legend. Right panel: Heliocentric radial velocities versus R.A. of the stars that remain after applying $\rm RV_{err} < 5~,km\,s^{-1}$. Colors are as in the left panel. The dashed horizontal lines show the used radial velocity selection criteria ($2\sigma$ clipping around the median).}
\label{fig:PMStars}
\end{figure*}

\begin{figure*}[h!]
\begin{subfigure}[h]{0.33\textwidth}
   \includegraphics[width=\textwidth]{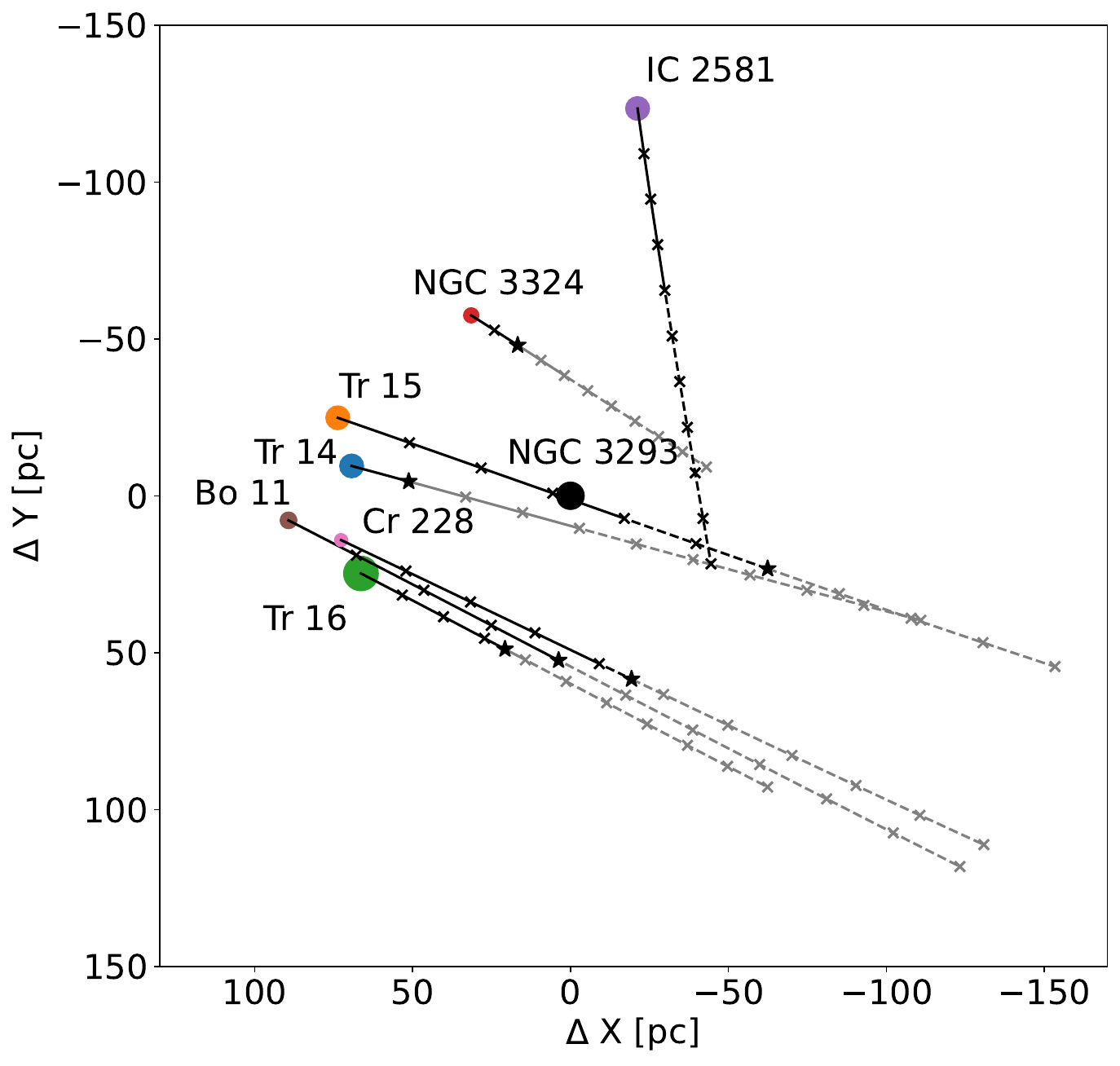}
\end{subfigure}
\begin{subfigure}[h]{0.33\textwidth}
   \includegraphics[width=\textwidth]{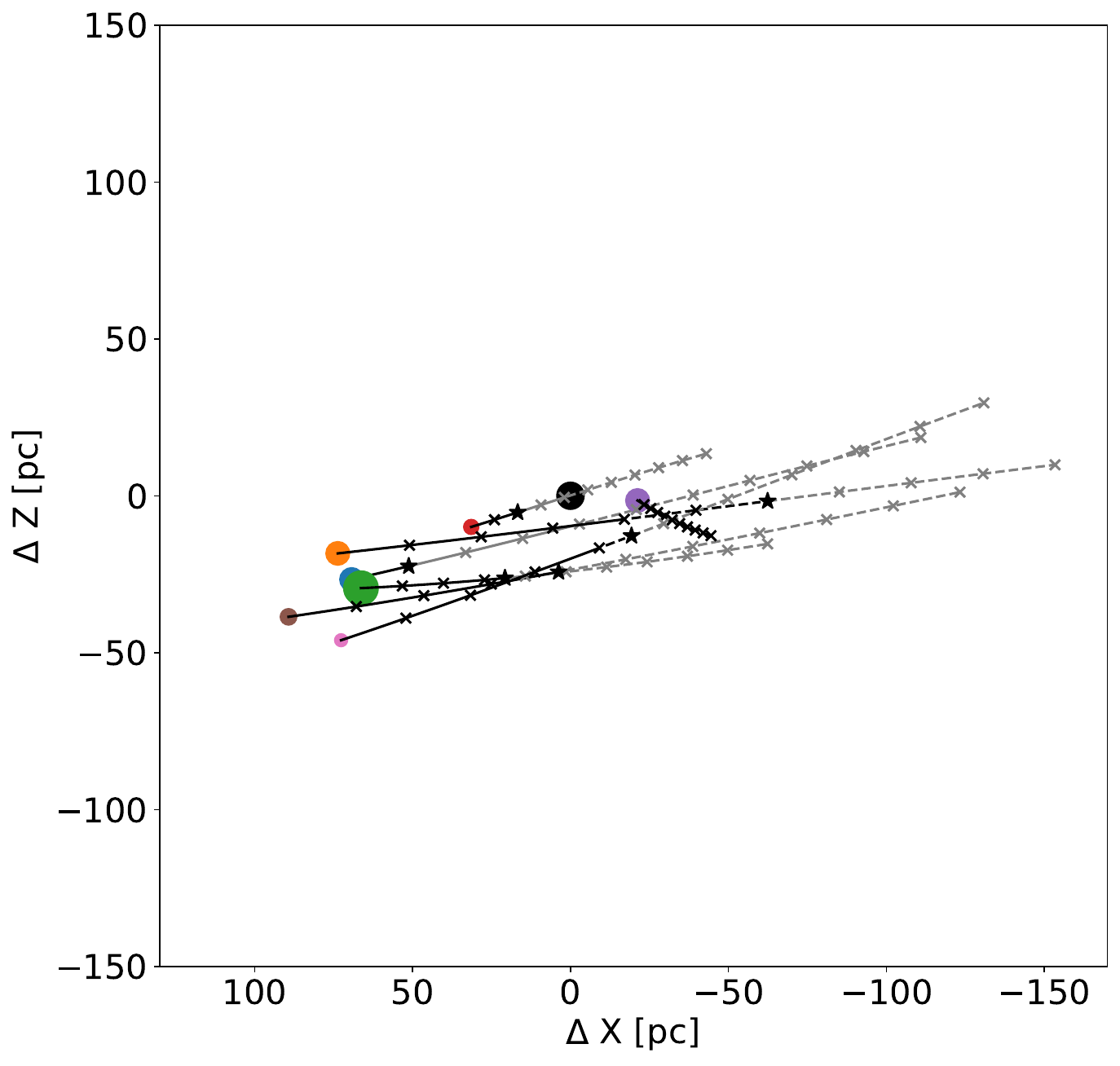}
\end{subfigure}
\begin{subfigure}[h]{0.34\textwidth}
   \includegraphics[width=\textwidth]{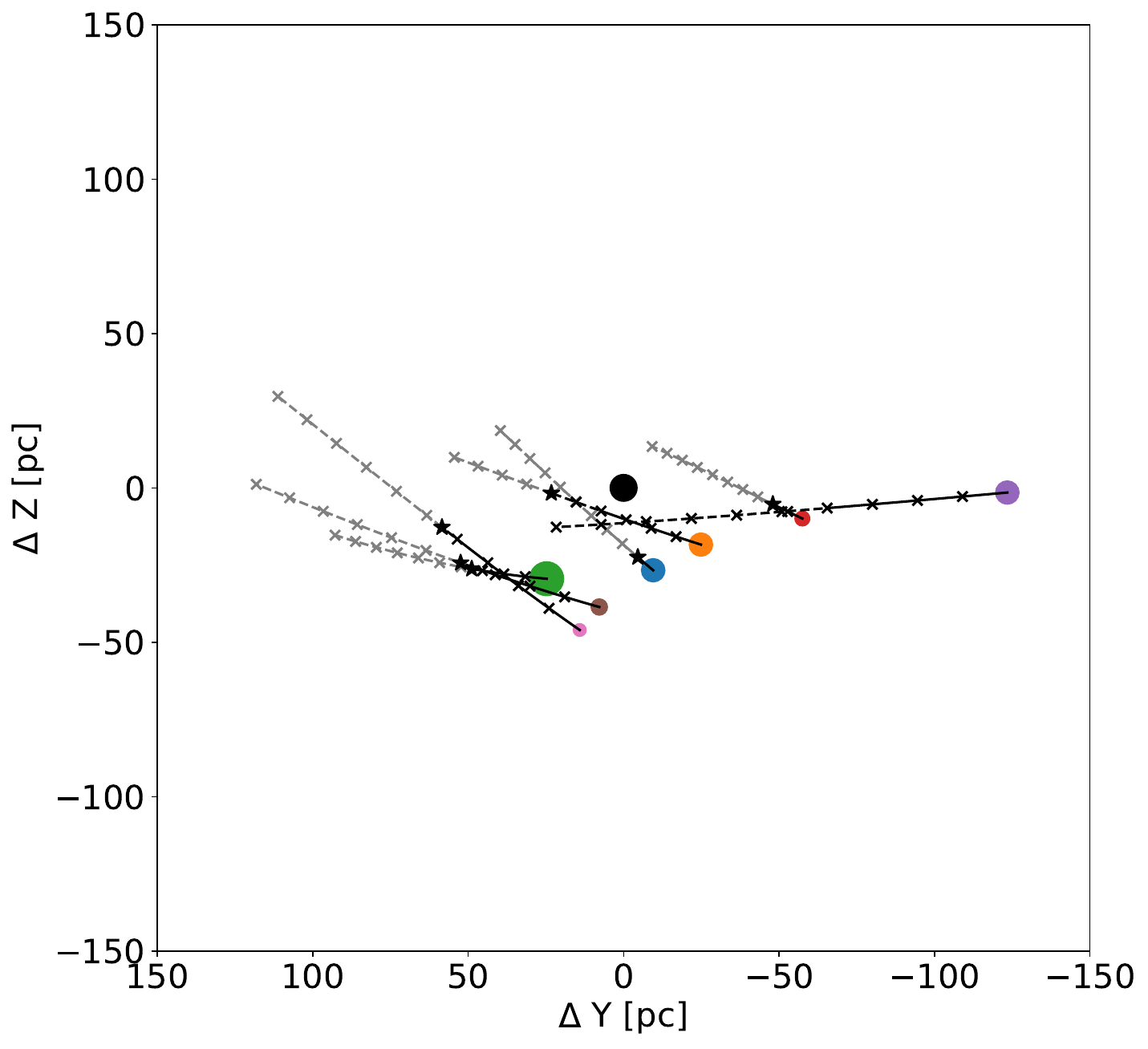}
\end{subfigure}
\caption{Paths of the clusters relative to NGC~3293 in the $(X,Y)$, $(X,Z)$ and the $(Y,Z)$ plane over the last 10~Myr. The paths before 4~Myr ago are plotted dotted, and the crosses show the cluster position for each megayear. The asterisk visualizes the time at which the cluster is formed.}
\label{fig:ClusterMotion-XY-XZ-YZ}
\end{figure*}

\begin{figure*}[h!]
\addtocounter{figure}{-1}
    \sidecaption
    \begin{subfigure}[h]{0.312\textwidth}
   \includegraphics[width=\textwidth, trim=0.cm 0.cm 3cm 0cm, clip]{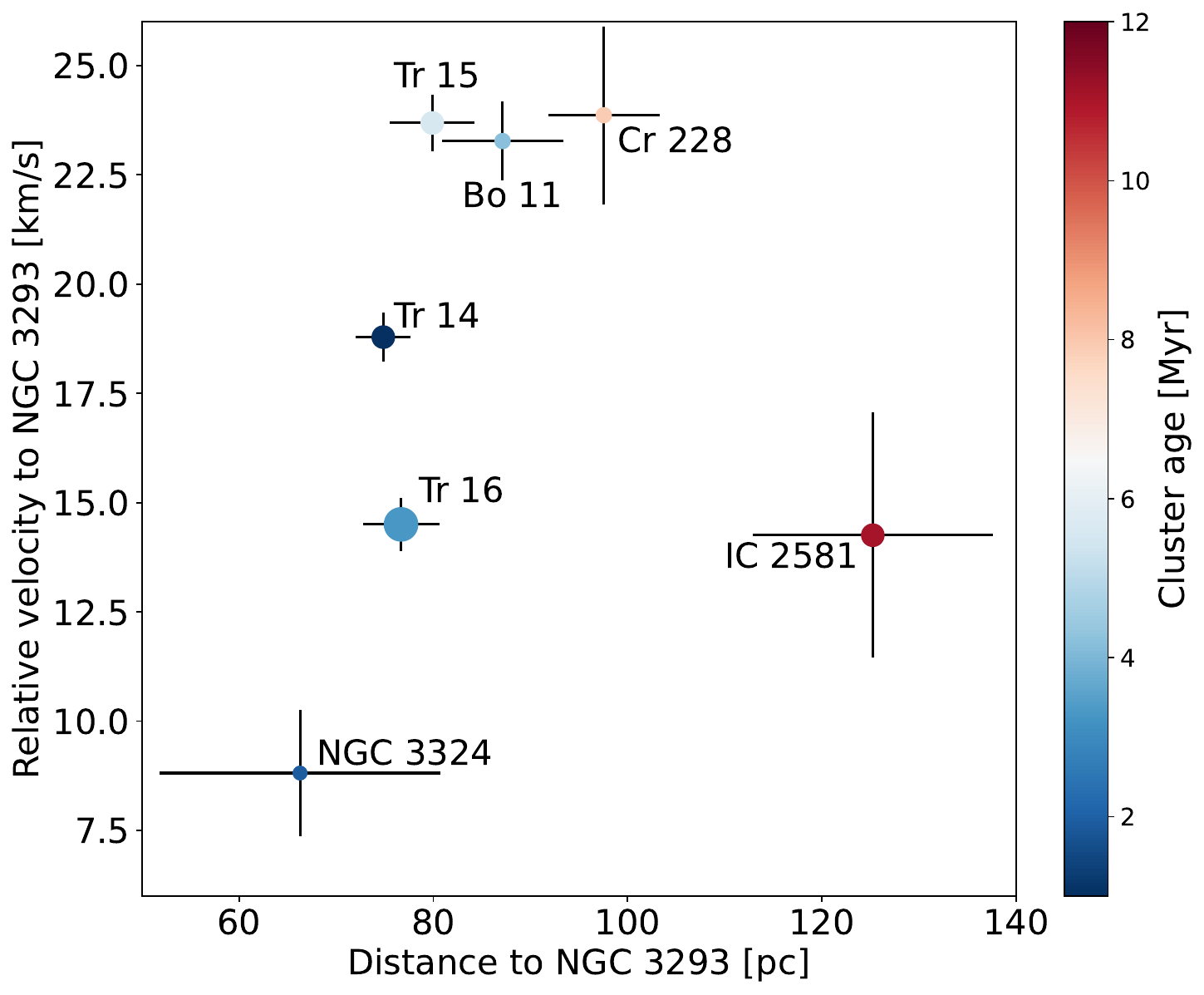}
\end{subfigure}
\begin{subfigure}[h]{0.348\textwidth}
   \includegraphics[width=\textwidth]{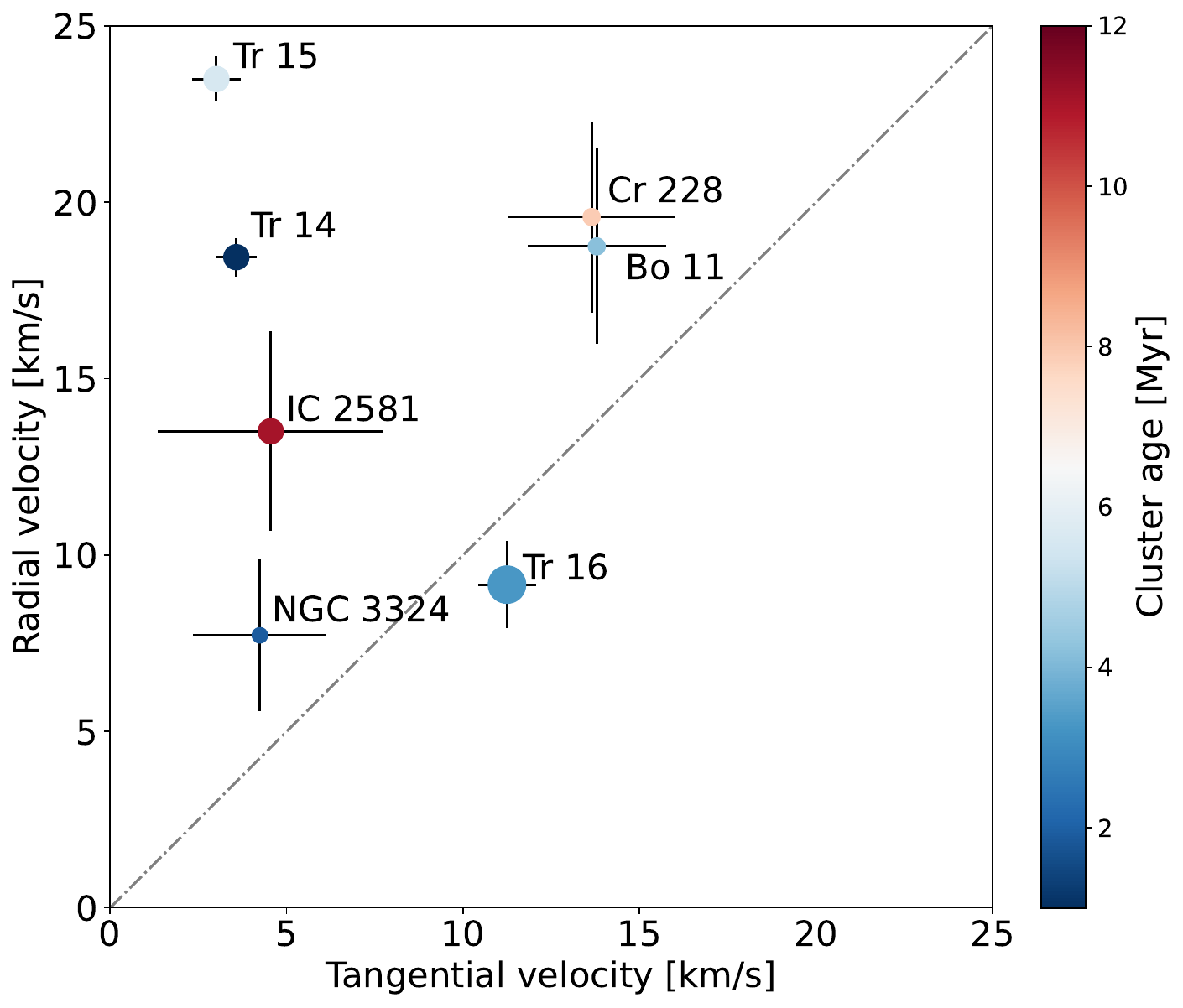}
\end{subfigure}
\caption{Left panel: Distances versus the relative velocity between NGC~3293 and the other clusters. Right panel: Tangential velocity component versus the radial velocity component of the clusters, with the gray line showing the 1-to-1 line.}
\label{fig:DisvsVel}
\end{figure*}

\begin{figure*}
\begin{minipage}{\columnwidth}
       
    \includegraphics[width=9cm]{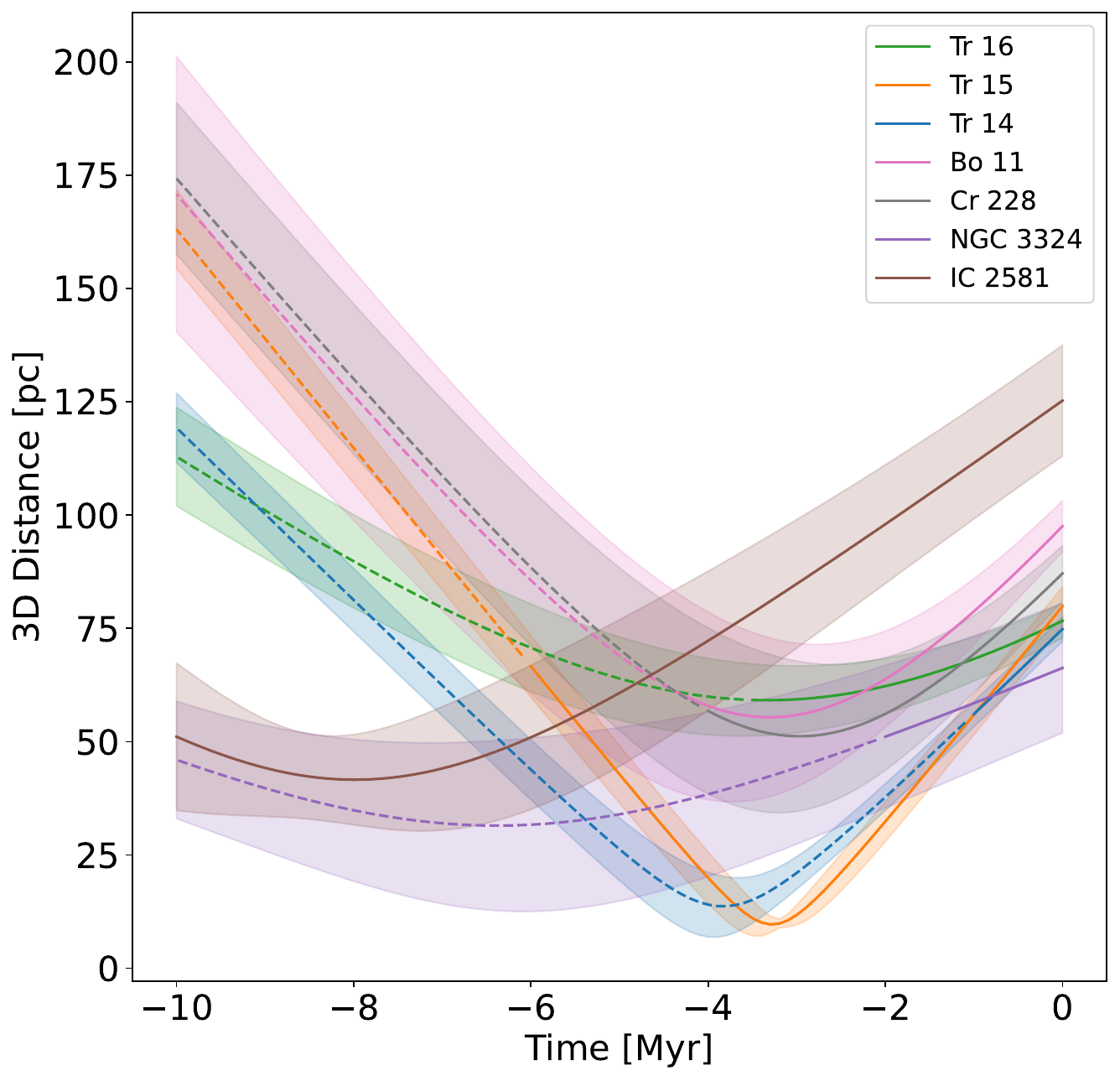}
    \caption{Distances between NGC~3293 and the other clusters over the last 10~Myr. Distances at times before the cluster was formed are shown as dashed lines. The shaded error ranges show the standard deviation around the mean distance.}
    \label{fig:ClusterDistances}

\end{minipage}
\end{figure*}

\begin{table*}
\caption{High-mass stars in our cluster sample.}
\resizebox{\hsize}{!}{\begin{tabular}{lllllccccccc} 
\hline\hline
Cluster & Gaia DR3 Designation & Name & Spectraltype & SpT Reference & RA & Dec & $\mu_\alpha^*~[\rm mas\,yr^{-1}]$ & $\mu_\delta~[\rm mas\,yr^{-1}]$ & $\varpi$ [mas] & RV $[\rm km\,s^{-1}]$ & RV Reference\\\hline
Tr 14 & Gaia DR3 5350383460949215232 & HD 93250 & O4 III((fc)) & 2014ApJS..211...10S& 10:44:45.01& $-59$:33:54.6& $-7.116$& 3.081& 0.424&&\\
Tr 14&Gaia DR3 5350386415886679424&Tr16-122&B1.5 V& 2016AJ....152..190A & 10:44:25.47&$-59$:33:09.4 & $-6.798$ & 2.095 & 0.442 & $-19.23$&1\\
Tr 14&Gaia DR3 5350362948184049280&Tr16-124&B1 V&2016AJ....152..190A&10:44:05.81&$-59$:35:11.7&$-6.596$&2.036&0.462&$-18.24$&1\\
Tr 14&Gaia DR3 5350362638946370688&Tr16-126&B0.7 V&2022A\&A...657A.131M&10:43:59.84&$-59$:35:24.3&$-6.715$&1.988&0.424&$-16.66$&1\\
Tr 14&Gaia DR3 5350363910256783488&HD 93129A&O2 If* + O2 If*&2014ApJS..211...10S&10:43:57.45&$-59$:32:51.4&$-6.277$&2.800&0.416&&\\
Tr 14&Gaia DR3 5350363910256783744&HD 93129B&O3.5 V((f))&2014ApJS..211...10S&10:43:57.63&$-59$:32:53.7&$-6.618$&2.182&0.419&&\\
Tr 14&Gaia DR3 5350363807162637696&HD 93128&O3.5 V((fc))z&2014ApJS..211...10S&10:43:54.38&$-59$:32:57.5&$-6.616$&2.135&0.437&&\\
Tr 14&Gaia DR3 5350363875896996480&CPD $-58$ 2611&O6 V((f))z&2014ApJS..211...10S&10:43:46.67&$-59$:32:54.8&$-6.747$&1.837&0.438&&\\
Tr 14&Gaia DR3 5350363944616553216&CPD $-58$ 2620&O7 V((f))z&2014ApJS..211...10S&10:43:59.89&$-59$:32:25.4&$-6.981$&1.642&0.373&&\\
Tr 14&Gaia DR3 5350363875897024256&Tr14-3&B0.2 V&2022A\&A...657A.131M&10:43:53.63&$-59$:33:00.7&$-6.450$&2.039&0.414&$-15.37$&1\\

\hline
\end{tabular}}
\tablefoot{The table contains the stars' spectral type and astrometric parameters. The table is available in its entirety in electronic form at the CDS.}
\label{tab:CarOB1HighMass}
\end{table*}

\begin{table*}
\caption{Galactocentric positions and velocities of the clusters over the last 13 Myr,
calculated with \textit{astropy}, where
the coordinate system is defined by the galactic center at $(X,Y) = (0,0)$ and the current solar position at $(X,Y) = (-8.23, 0)$\;kpc.} 
\resizebox{\hsize}{!}{\begin{tabular}{clcccccc} 
\hline\hline
Time [Myr] & Cluster & $\rm X + \sigma_X$ [kpc] & $\rm Y + \sigma_Y$ [kpc] & $\rm Z + \sigma_Z$ [kpc] & $\rm U + \sigma_U~[km\,s^{-1}]$ & $\rm V + \sigma_V~[km\,s^{-1}]$ & $\rm W + \sigma_W~[km\,s^{-1}]$ \\\hline

$\;\;0.0$&Tr 14&$-7.525\pm0.002$&$-2.246\pm0.006$&$-0.008\pm0.001$&$-63.348\pm0.356$&$216.587\pm0.925$&$-5.961\pm0.042$\\
$-0.1$&Tr 14&$-7.519\pm0.002$&$-2.268\pm0.006$&$-0.008\pm0.001$&$-63.957\pm0.356$&$216.404\pm0.925$&$-5.965\pm0.042$\\
$-0.2$&Tr 14&$-7.512\pm0.002$&$-2.290\pm0.006$&$-0.007\pm0.001$&$-64.566\pm0.356$&$216.219\pm0.925$&$-5.970\pm0.042$\\
$-0.3$&Tr 14&$-7.506\pm0.002$&$-2.312\pm0.006$&$-0.006\pm0.001$&$-65.174\pm0.356$&$216.033\pm0.925$&$-5.974\pm0.042$\\
$-0.4$&Tr 14&$-7.499\pm0.002$&$-2.334\pm0.006$&$-0.006\pm0.001$&$-65.782\pm0.355$&$215.845\pm0.925$&$-5.978\pm0.042$\\
$-0.5$&Tr 14&$-7.492\pm0.002$&$-2.356\pm0.006$&$-0.005\pm0.001$&$-66.389\pm0.355$&$215.655\pm0.925$&$-5.981\pm0.042$\\
$-0.6$&Tr 14&$-7.485\pm0.002$&$-2.378\pm0.006$&$-0.005\pm0.001$&$-66.995\pm0.355$&$215.463\pm0.925$&$-5.984\pm0.042$\\
$-0.7$&Tr 14&$-7.479\pm0.002$&$-2.400\pm0.006$&$-0.004\pm0.001$&$-67.601\pm0.355$&$215.268\pm0.925$&$-5.986\pm0.042$\\
$-0.8$&Tr 14&$-7.472\pm0.002$&$-2.422\pm0.006$&$-0.003\pm0.001$&$-68.206\pm0.355$&$215.075\pm0.925$&$-5.989\pm0.042$\\
$-0.9$&Tr 14&$-7.465\pm0.002$&$-2.444\pm0.006$&$-0.003\pm0.001$&$-68.811\pm0.355$&$214.877\pm0.925$&$-5.991\pm0.042$\\

\hline
\end{tabular}}
\tablefoot{The table is available in its entirety in electronic form at the CDS.}
\label{tab:ClusterTraceback}
\end{table*}

\end{appendix}

\end{document}